\documentclass{aa}

\usepackage{graphicx}
\usepackage{natbib}
\usepackage{scalerel}
\usepackage{amsmath}	% Advanced maths commands
\usepackage{multicol}        % Multi-column entries in tables
\usepackage{bm}		% Bold maths symbols, including upright Greek
\usepackage{pdflscape}	% Landscape pages
\usepackage[usenames,dvipsnames,table]{xcolor}
\usepackage{tabularx}
\usepackage{multirow}
\usepackage{multicol}
\usepackage{tablefootnote}
\usepackage{threeparttable}
\usepackage{comment}

\newcommand{\camels}{$\textsc{camels}$}

\usepackage{txfonts}
\usepackage[pdfencoding=auto,psdextra]{hyperref}
\hypersetup{
    colorlinks=true,
    linkcolor=blue,
    filecolor=magenta,      
    urlcolor=blue,
    citecolor=blue
}
\makeatletter
\renewcommand*\aa@pageof{, page \thepage{} of \pageref*{LastPage}}
\makeatother

\usepackage[utf8]{inputenc}

\usepackage[switch, modulo]{lineno}
\def\Msun{\mbox{M$_\odot$}}

\def\ML{\mbox{$M/L$}}

\def\mst{\mbox{$M_{\star}$}}

\def\lsim{\mathrel{\rlap{\lower3.5pt\hbox{\hskip0.5pt$\sim$}}
    \raise0.5pt\hbox{$<$}}}                % less than or approx. symbol
\def\gsim{~\rlap{$>$}{\lower 1.0ex\hbox{$\sim$}}}

\def\Fig{\mbox{Fig.~}}
\def\Figs{\mbox{Figs.~}}
\def\Tab{\mbox{Table~}}

\def\Sec{\mbox{Section~}}

\def\App{\mbox{Appendix~}}

\begin{document}
%
% Put the title and authors of your paper here
%

\title{CASCO: Cosmological and AStrophysical parameters from Cosmological simulations and Observations}
\subtitle{III. The physics behind the emergence of the golden mass scale}

%% please do not edit the author list once you copy it from the
   
\newcommand{\orcid}[1]{} %% define as link to https://orcid.org/#1 if needed

\author{
C. Tortora,$^{1}$\thanks{E-mail: crescenzo.tortora@inaf.it}
V. Busillo,$^{1,2,3}$
N. R. Napolitano,$^{2}$
L. V. E. Koopmans,$^{4}$
G. Covone,$^{2,3}$
S. Genel,$^{5,6}$
F. Villaescusa-Navarro,$^{5,7}$
M. Silvestrini$^{1,2}$} 

\institute{
$^{1}$INAF -- Osservatorio Astronomico di Capodimonte, Salita Moiariello 16, I-80131, Napoli, Italy\\
$^{2}$Dipartimento di Fisica “E. Pancini”, Universit\`{a} degli studi di Napoli Federico II, Compl. Univ. di Monte S. Angelo, Via Cintia, I-80126 Napoli, Italy\\
$^{3}$INFN, Sez. di Napoli, Compl. Univ. di Monte S. Angelo, Via Cintia, I-80126 Napoli, Italy\\
$^{4}$Kapteyn Astronomical Institute, University of Groningen, P.O Box 800, 9700 AV Groningen, The Netherlands\\
$^{5}$ Center for Computational Astrophysics, Flatiron Institute, 162 5th Avenue, New York, NY 10010, USA\\
$^{6}$ Columbia Astrophysics Laboratory, Columbia University, 550 West 120th Street, New York, NY 10027, USA\\
$^{7}$ Department of Astrophysical Sciences, Princeton University, 4 Ivy Lane, Princeton, NJ 08544 USA\\}

% 
% Put your abstract here:
%

\abstract{Observations reveal a characteristic ``golden mass''---around $10^{12} \, \Msun$ in halo mass and $5 \times 10^{10} \, \Msun$ in stellar mass---associated with a peak in star formation efficiency. Using \textsc{camels} simulations based on IllustrisTNG in a $(50\, \mathrm{h^{-1}}\,\mathrm{Mpc})^3$ volume, we investigate how this scale arises and evolves under varying supernova (SN) and active galactic nucleus (AGN) feedback strengths and cosmological parameters ($\Omega_{\rm m}, \sigma_8$). We find a U-shaped relation between dark-to-stellar mass ratio (within the half-mass radius) and stellar mass, with a minimum at the golden mass, in line with observations. Cosmology primarily shifts the normalization of the scaling relation, while SN and AGN feedback modify both the shape and the emergence of the golden mass. Stronger SN feedback shifts the golden mass to lower values, while AGN feedback---especially radiative efficiency (i.e., the fraction of the accretion rest mass released in the accretion process), followed by the black hole feedback factor (i.e., the normalization factor for the energy in the AGN feedback in the high-accretion state) and quasar threshold (i.e. the Eddington ratio)---affects the high-mass slope and shifts the golden mass value. The golden mass is appearing earlier in cosmic time for simulations with stronger feedback, which more rapidly quench star formation in massive galaxies. Splitting galaxies by star formation activity reveals that passive galaxies preserve the U-shape, while star-forming ones show a decreasing dark matter fraction with stellar mass, with hints of a reversal at low redshift. Global stellar fractions also follow a U-shaped trend. However, in passive systems, the golden mass disappears, shifting to lower masses, while star-forming galaxies exhibit a peak only at low redshift. Our results highlight feedback as the primary driver behind the emergence of the golden mass up to $z \sim 1.5 - 2$, while stream and virial shock processes play a secondary role. Comparing our results with other theoretical expectations and observational findings, we speculate that at $z \gtrsim 1.5-2$, a single characteristic (stream) mass regulates galaxy evolution, which later bifurcates into two: a low-mass ``gas-richness" scale tied to gas availability, and a higher-mass golden mass governing star formation efficiency and quenching.}

%
% Provide up to five key words:
%
%    from the list in
%     https://www.aanda.org/for-authors/latex-issues/information-files#pop}
%
% Add short versions of title and author list for page headings
%
   \titlerunning{CASCO-III}
   \authorrunning{C. Tortora}
   
   \maketitle
%
%-------------------------------------------------------------------
%
%
%   Start the main text of your paper here

\section{Introduction}

Research using spectroscopic and photometric data have led to the characterization of scaling relations among various galaxy parameters, including stellar populations (age, metallicity, luminosity, stellar mass, IMF slope), structural properties (such as galaxy size and light profile), and total mass and dark matter (DM) distribution (DM fraction, mass density slope). Understanding the origins of these scaling relations, the physical processes shaping them and their role in galaxy evolution across a vast range of masses and cosmic time is central to modern studies of galaxy evolution \citep[e.g.,][]{Courteau+14_review,DOnofrio+21_review}.

There is growing evidence that a critical stellar mass scale exists around $\sim 5 \times 10^{10} \, \rm \Msun$, corresponding to $\sim 10^{12} \, \rm \Msun$ in virial mass \citep[e.g.][]{dek_birn06}. This characteristic mass, often referred to as the "golden mass" (or "transition mass" or "bimodality mass"), appears as transition, break, or extremum in the trends of various scaling relations. Given the power-law nature of the DM power spectrum from simulations \citep{Widrow+09}, there is no reason to consider any particular mass scale as special. In contrast, the stellar mass function follows a Schechter form. As a result, near the break of the stellar mass function, star formation efficiency, i.e. the ratio of stellar to halo mass normalized by the cosmological baryon fraction ($\epsilon_{\rm SF} = \mst/(M_{\rm h} f_{\rm b})$), reaches its peak, whereas at both higher and lower mass scales, feedback mechanisms reduce this efficiency. Thus, any characteristic mass scale is likely tied to baryonic processes rather than DM dynamics.

If this is indeed a fundamental mass scale in galaxy formation and evolution, it is plausible that the physical processes driving galaxy evolution also change when crossing this mass threshold. One well-studied scaling relation involves total virial mass and stellar mass, which can be translated into a relation between the star formation efficiency and stellar or virial mass (e.g. \citealt{Moster+10}). The bending observed in the virial-to-stellar mass relation translates into a peak in the correlation between stellar fraction (or star formation efficiency) and mass at the golden mass. Analyzing the correlation between stellar fraction or star formation efficiency with mass can provide insights into the role of various galactic processes, for example SN and AGN feedback, which are believed to dominate at low and high masses, respectively. At the golden mass, supernova-driven winds and AGN feedback become less effective at depleting the galaxy's gas supply and quenching star formation (e.g., \citealt{dek_birn06, Moster+10}). Moreover, \cite{dek_birn06} have proposed that one of the primary driver of the bimodality in galaxy properties is the transition from cold gas inflows to virial shock heating at a critical virial mass scale of $\sim 10^{12}\, \rm \Msun$, in conjunction with cold streams at high redshifts, feedback processes and gravitational clustering \citep{Daddi+22_main_sequence,Daddi+22_cold_hot}.

DM fraction within the half-light radius in early-type systems exhibit a U-shaped trend (\citealt{TLBN16_IMF_dwarfs, Tortora+19_LTGs_DM_and_slopes}), resembling the trend of star-formation efficiency on a global scale. However, tensions arise when considering different galaxy types. A bending in the trend of star formation efficiency with mass is not seen in recent analyses of rotation curves of star-forming galaxies (\citealt{Posti+19}).  Using the same data as in \cite{Posti+19}, \cite{Tortora+19_LTGs_DM_and_slopes} only found some mild evidence of a bending in such scaling relations.

Understanding the nature of the golden mass is therefore a central question to unravelling the physical processes governing galaxy formation. Despite progress, the physics underlying these scaling relations remains unclear, leaving key questions: What processes shape the observed scaling relations, and to what extent? Why does the golden mass emerge, and how does it manifest? Is this golden mass constant in time or does it show up at some specific epoch as a consequence of a dominating physical mechanism governing its appearance? How do these relations and the golden mass vary with  galaxy types? Are different extrema and bends in scaling relations all measuring the same mass scale? In particular, it would be interesting to analyze whether feedback, virial shocks and cold streams act independently, are interconnected, or if one plays a more dominant role than the others in explaining the golden mass. All these questions can be addressed by exploring the predictions of cosmological simulations and comparing them with observations (e.g. \citealt{Mukherjee2018,Mukherjee2021,Mukherjee2022}; \citealt{Busillo2023,Busillo+24_CASCO-II}).

We began exploring these issues in the first two papers of the "Cosmological and AStrophysical parameters from Cosmological simulations and Observations" (CASCO) series, using \camels\ simulations within a volume of $(25 \, h^{-1}\,\textrm{Mpc})^{3}$, one of the most advanced suites of cosmological simulations, which allows for variation in cosmological parameter values and, most relevantly for this paper, the strength of SN and AGN feedback \citep{Villaescusa-Navarro2021, Villaescusa-Navarro2022,Ni+23}. By identifying the best simulations that reproduce different observed scaling relations among central DM mass and fraction, total virial mass, and half-mass radius versus stellar mass for different data samples (for both passive and star-forming galaxies) in \citet[Paper I hereafter]{Busillo2023} and \citet[Paper II hereafter]{Busillo+24_CASCO-II}, we have constrained cosmological and astrophysical parameters.

In this third paper of the CASCO series, we address the aforementioned open questions from a theoretical point of view, focusing on the origin of the golden mass, using the latest \camels\ simulations, which improve the galaxy count statistics by increasing the simulation volume eightfold to $(50\, h^{-1}\,\textrm{Mpc})^{3}$. We examine the scaling relations between the total-to-stellar mass ratio within the half-mass radius and stellar mass, as well as the correlation between half-mass radius and stellar mass, and total stellar mass fraction and stellar mass, as a function of redshift to decipher the impact of SN and AGN feedback.

The paper is organized as follows. We start providing a summary of the observational and simulation findings on the golden mass in \Sec\ref{sec:state-of-the-art}. In \Sec\ref{sec:Data}, we present the simulations. Results on the effect of astrophysical and cosmological parameters on the scaling relations and the golden mass are discussed in \Sec\ref{sec:golden_mass}. The results are discussed in a broader context and compared with the literature in \Sec\ref{sec:discussion} and a summary of paper findings with future prospects is provided in \Sec\ref{sec:conclusions}.

\section{The state of the art on the golden mass}\label{sec:state-of-the-art}

\subsection{Characteristic mass scales in galaxy evolution}

Across the galaxy mass spectrum characteristic transition masses in galaxy properties exist. In particular, \cite{Kannappan+13} has identified two characteristic mass scales in galaxy morphology, gas fractions, and fueling regimes (see also \citealt{Hunt2020}), which separate three regimes: "accretion-dominated" dwarf galaxies, "gas-equilibrium" or "processing-dominated" intermediate-mass galaxies and, "gas-poor" or "quenched" massive galaxies. Below a stellar mass threshold of $\sim 3 \times 10^{9}\, \rm \Msun$ dwarf galaxies are characterized by an overwhelming gas accretion, this mass scale can appear as a break in the HI-mass vs stellar mass relation \citep{Hunt2020}. The mass scale of  $\sim 3 \times 10^{9}\, \rm \Msun$ is termed "gas-richness" threshold. Above this mass we enter into a "processing-dominated" regime, where such intermediate-mass galaxies are able to consume gas through star formation as fast as it
is accreted. This regime ends up at a mass threshold of $\sim 3 \times 10^{10}\, \rm \Msun$, the bimodality or golden mass. Above this mass scale, we have a "gas-poor" or "quenched" regime, populated by spheroid-dominated, gas-poor galaxies with low rates
of star formation over the last Gyr.

\subsection{The golden mass from observations}

The "golden mass" is observed in several contexts, including the trends of total mass-to-light ratio and halo mass with stellar mass (e.g., \citealt{Benson+00,MH02,vdB+07,CW09,Moster+10, Moster2013,Behroozi+13,Rodriguez-Puebla+15,Behroozi+19,Girelli+20}). The stellar-to-halo mass relation (SHMR), or equivalently the star formation efficiency–mass relation, has been extensively studied in recent years, both in cumulative form and by separating galaxy samples by type. The bending in the SHMR translates into a peak in the trend between star formation efficiency and mass. This value is  $M_{\rm h} \sim 10^{12} \, \rm \Msun$ when DM halo masses are considered, corresponding to a stellar mass value of $\sim 3 \times 10^{10} \, \rm \Msun$.

It also manifests in the half-light dynamical \ML\ and DM fraction (\citealt{Wolf+10, Toloba+11_I, Cappellari+13_ATLAS3D_XX, TLBN16_IMF_dwarfs,Tortora+19_LTGs_DM_and_slopes, Aquino-Ortiz+18,Lovell+18_Illustris,Busillo2023,Busillo+24_CASCO-II}), in the gradients of dynamical \ML\ profiles across several effective radii (\citealt{Napolitano+05}), and in the total mass density slope of galaxies (\citealt{Tortora+19_LTGs_DM_and_slopes}). Historically, this mass scale has emerged from structural analyses, such as the correlation between effective surface brightness and effective radius  (\citealt{Capaccioli+92a}, \citealt{TullyVerheijen97}, \citealt{Kormendy+09}) and size-mass relations (\citealt{Shen+03, HB09_curv, Tortora+09,Nedkova+21}), as well as in trends of optical color, metallicity, and stellar \ML\ gradients (\citealt{Kuntschner+10, Spolaor+10, Tortora+10CG, Tortora+11MtoLgrad}). Different works also find a bending in the main sequence of star-forming galaxies, which is associated to the golden mass \citep{Daddi+22_main_sequence,Popesso+23,Stephenson+24,Enia+25_Euclid}. This extensive body of empirical evidence mostly converges on a value for the golden stellar mass around $1 \, \textrm{-} \, 5 \times 10^{10}\, \Msun$ at $z \sim 0$.

Tensions arise when the SHMR is analyzed considering different galaxy types (e.g., \citealt{Dutton+10,More+11,Wojtak_Mamon13,Posti+19,Correa_Schaye2020}). A bending in the SHMR is not seen in the analyses of rotation curves of star-forming galaxies by \cite{Posti+19}, pointing to larger star formation efficiencies of late-type systems when compared to early-types of similar stellar mass above the golden mass. Building SHMRs through statistical approaches that combine various semi-empirical methods of galaxy–halo connection, \cite{Rodriguez-Puebla+15} found a segregation in color for central galaxies, with bluer galaxies having a larger stellar fraction than red galaxies at fixed virial mass. Using SDSS data with visual classifications, \cite{Correa_Schaye2020} found that disks have stellar masses (and hence star formation efficiencies) that are larger or equal to those of ellipticals at virial masses below $10^{13}\, \rm \Msun$, and less massive at larger virial masses; EAGLE simulations also indicate that disks are more massive than ellipticals at fixed virial mass. In contrast to \cite{Posti+19}, both studies observed the golden mass in both galaxy types. When the same quantities are calculated in the central galactic regions and transformed into a DM fraction within the half-light radius, early-type galaxies exhibit a U-shaped trend (\citealt{TLBN16_IMF_dwarfs, Tortora+19_LTGs_DM_and_slopes}). Using the same data as in \cite{Posti+19}, \cite{Tortora+19_LTGs_DM_and_slopes} only found some mild evidence of a bending in such scaling relations for late-type galaxies.

Some analysis, mostly concerning the SHMR and the main-sequence of star-forming galaxies have probed bendings as a function of redshift. \cite{Behroozi+13} shows that the instantaneous star formation efficiency peaks at $M_{\rm h} \sim 10^{11.7}\, \rm \Msun$ at $z\sim 0$, and varies very little since $z=4$. Modelling the main sequence of star-forming galaxies, using a power-law with a turnover mass, \cite{Popesso+23} find a value, in stellar mass, of $\sim 10^{10} \, \rm \Msun$ at low redshifts, which increases to $\sim 10^{11}\, \rm \Msun$ at $z \sim 6$. Translated to halo masses, these authors find a mass of $\sim 10^{11.4-11.5}\, \rm \Msun$ at $z \sim 0$, with a steep increase at $z \gsim \, 0.8$, up to $\sim 10^{13}\, \rm \Msun$ at $z \sim 6$ \citep[see also][]{Lee+15,delvecchio+21,Daddi+22_main_sequence,Enia+25_Euclid}.  
Analyzing the size-mass relation as a function of redshift, \cite{Nedkova+21} find a monotonic trend in star-forming galaxies, while the relation is steeper for quiescent galaxies with
stellar masses above $\sim 10^{10.3} \, \rm \Msun$ and flattens at lower masses. From their Figure 10 we observe that the bending mass for passive galaxies is increasing with redshift.

\subsection{Theoretical interpretation of the golden mass}

If historically SN and AGN feedback have been supposed to be the leading processes in galaxies evolution, with a varying role as a function of mass, more recently the role of the virial shock heating and cold streams have emerged as alternative or complementary processes. The interplay of such processes is supposed to influence the emergence of the golden mass.

In galaxies with stellar masses below the golden mass, SN feedback plays a crucial role in regulating star formation. The energy released by supernova explosions can heat the surrounding circum-galactic medium (CGM) and, in some cases, expel gas from the galaxy. This process is particularly effective in low-mass galaxies, where the gravitational potential wells are shallower, making it easier for the gas to escape. The effectiveness of SN feedback decreases with increasing galaxy mass, as the deeper gravitational potentials in more massive galaxies can better retain the heated gas. Using the theory of supernova bubble in \cite{Dekel_Silk86}, equating the energy released in the interstellar medium to the the binding energy of this gas in the DM halo potential well, gives an upper limit for the mass of a halo in which
SN feedback can be effective of $M_{\rm SN} \sim 10^{11.7}\, \rm \Msun$ at $z\sim 0$, scaling with redshift as $(1+z)^{-3/2}$ \citep{dek_birn06}.

At low masses, SN feedback suppresses the supermassive black holes (SMBHs) growth. In more massive haloes, the deep gravitational potential and hot CGM confine the central gas, enabling sustained accretion onto the black hole \citep{Dubois+15,Bower+17_EAGLE,Habouzit+19_IllustrisTNG}. Therefore, above a critical mass scale
the growth of SMBHs at the centers of galaxies becomes more pronounced. The energy and momentum output from AGN can drive powerful outflows, heating and expelling gas from the galaxy and its surrounding CGM. This AGN feedback further suppresses star formation by depleting the cold gas reservoir necessary for star formation and maintaining the hot state of the CGM. The onset of significant AGN activity around the golden mass is also associated with morphological transformations in galaxies, such as the transition from disk-like to spheroidal structures. From these considerations the maximum mass for effective SN feedback emerge as a lower limit for impactful AGN feedback ($M_{\rm AGN} \sim M_{\rm SN}$).

Virial shock heating and cold streams are competing, complementary or supporting processes \citep{dek_birn06,Dekel+19,Stern+20,Aung+24,Waterval+25}. Hydrodynamical simulations predict that for low halo masses gas accretes in cold filaments (“cold-mode accretion”) directly to the galaxy disk, efficiently forming stars. However, as galaxies grow in mass, they reach a point where the infalling gas is shock-heated to the virial temperature of the DM halo (“hot-mode accretion”) and star formation is rapidly quenched. This virial shock heating becomes significant in halos with masses around $M_{\rm shock} \sim 10^{11.7}\, \rm \Msun$ (which is predicted to be redshift-independent for $z<3$), which approximately corresponds to the mass where SN and AGN feedback efficiency is minimal. The shock-heated gas forms a hot, diffuse CGM that is less prone to cooling and condensing into the galaxy, thereby suppressing further star formation. This mechanism is supposed to contribute to the decline in star formation efficiency observed in more massive galaxies. However, at $z \gtrsim 1.5$, narrow cold streams are expected to penetrate the otherwise hot CGM and sustain efficient star formation, even in haloes exceeding the critical mass for shock heating (e.g. \citealt{Keres+05}).

Recent works show that the bending in the star-forming main sequence across redshift can be connected to virial shocks and cold streams \citep{Daddi+22_main_sequence,Popesso+23,Stephenson+24}. In particular, when converting turnover mass in the main sequence of star-forming galaxies into hosting DM halo masses, \cite{Daddi+22_main_sequence} find consistent results with the evolving cold- to hot-accretion transition mass, defined by the redshift-independent
$M_{\rm shock}$ at $z < 1.4$ and by the rising $M_{\rm stream}$ at $z > 1.4$.

Given the considerations above, several key aspects remain unexplored. It is still unclear whether feedback processes, virial shocks, and the physics of the CGM represent competing scenarios, or whether they offer independent or interconnected explanations for the origin of the golden mass. The \camels\ cosmological simulations can help disentangle these effects by assessing the impact of feedback mechanisms and quantifying the role of virial shocks when feedback is suppressed.

Although less explored in the literature, we will center our analysis on the relationship between central DM content and stellar mass. These are measurable quantities that are expected to become increasingly accessible in the near future through dynamics using spectroscopic surveys \citep{deJongR+11_4MOST,Iovino+23_StePS-WEAVE,Iovino+23_StePS-4MOST} and strong lensing \citep{Walmsley+25_SLDE_A}. Our goal is to derive theoretical expectations for the golden mass based on this correlation and compare them with independent observational findings.

\section{Data}\label{sec:Data}

\begin{table*}
\centering
\caption{List of free parameters of \camels\ simulations used in this paper. See more details in \cite{Ni+23}.}\label{tab:parameters}
\resizebox{\textwidth}{!}{%
\begin{tabular}{cl}
\hline
\rm Parameter &  \rm Description\\
\hline
$\Omega_{\rm m}$ & cosmological mass density parameter\\
$\sigma_{8}$ &  rms of the $z=0$ linear overdensity in spheres of radius $8 \, \rm h^{-1}$ Mpc\\
\hline
$A_{\rm SN1}$ & normalization
factor for the energy in galactic winds per unit star
formation\\
$A_{\rm SN2}$ & normalization
factor for the galactic wind speed\\
\hline
$A_{\rm AGN1}$ & radio feedback factor, i.e., normalization
factor for the energy in AGN feedback, per unit accretion
rate, in the low-accretion state\\
$A_{\rm AGN2}$ & radio feedback reiorientation factor, i.e., normalization factor for the frequency of
AGN feedback energy release events\\ 
$BH_{\rm accr}$ & BH accretion factor, i.e., normalization
factor for the Bondi rate for the accretion onto BHs\\
$BH_{\rm Edd}$ & BH Eddington factor, i.e., normalization
factor for the limiting Eddington rate for the accretion
onto BHs\\
$BH_{\rm FF}$ & BH feedback factor, i.e., normalization
factor for the energy in AGN feedback, per unit accretion
rate, in the high-accretion state\\
$BH_{\rm RE}$ & BH radiative efficiency, i.e., the fraction of the
accretion rest mass that is released in the accretion process\\
$Q_{\rm T}$ & Quasar threshold, i.e., the Eddington ratio, that serves as the threshold
between the low-accretion and high-accretion states of
AGN feedback\\
$Q_{\rm TP}$ & Quasar threshold power, i.e., the power-law index of the scaling relation of the low- to high-accretion state threshold with BH mass\\
\hline
\end{tabular}}
\end{table*}

\begin{table}
\centering
\resizebox{\columnwidth}{!}{%
\begin{threeparttable}
\caption{Values of the '1P' simulation parameters listed in \Tab\ref{tab:parameters}.}\label{tab:params_values}
%\resizebox{\columnwidth}{!}{%
\begin{tabular}{cccccc}
\hline
\rm Parameter & \multicolumn{5}{c}{Variations}\\
\rm & n2 & n1 & fiducial & p1 & p2  \\ 
\hline
$\Omega_{\rm m}$ & 0.1 & 0.2 & 0.3 & 0.4 & 0.5  \\
$\sigma_{8}$ & 0.6 & 0.7 & 0.8 & 0.9 & 1 \\
\hline
$A_{\rm SN1}$ &  0.9 & 1.8 & 3.6 & 7.2 & 14.4 \\
$A_{\rm SN2}$ & 3.7 & 5.2 & 7.4 & 10.5 & 14.8  \\
\hline
$A_{\rm AGN1}$ & 0.25 & 0.5 & 1 & 2 & 4 \\
$A_{\rm AGN2}$ & 10 & 14.1 & 20 & 28.3 & 40  \\
$BH_{\rm accr}$ & 0.25 & 0.5 & 1 & 2 & 4  \\
$BH_{\rm Edd}$ & 0.1 & 0.32 & 1 & 3.2 & 10  \\
$BH_{\rm FF}$ & 0.025 & 0.05 & 0.1 & 0.2 & 0.4  \\
$BH_{\rm RE}$ & 0.05 & 0.1 & 0.2 & 0.4 & 0.8   \\
$Q_{\rm T}$ & $6.33 \times 10^{-5}$ & $3.6 \times 10^{-4}$ & 0.002 & 0.011 & 0.063 \\
$Q_{\rm TP}$ & 0 & 1 & 2 & 3 & 4  \\
\hline
\end{tabular}
\begin{tablenotes}
        \footnotesize
        \item $^1$ The parameter values are labeled as n2, n1, fiducial, p1, and p2, ranging from the lowest to the highest. In the figures, for the astrophysical parameters, values are normalized relative to the fiducial simulation.
        \item $^1$ Each row in the table corresponds to a 1P simulation set, where the parameter in the first column is varied while the others remain fixed.
    \end{tablenotes}
\end{threeparttable}
}
\end{table}

Following Papers I and II, we use the simulated galaxy data coming from \camels, a suite of cosmological simulations, originally within a cosmological volume of $(25\,h^{-1}\;\textrm{Mpc})^{3}$ \citep{Villaescusa-Navarro2021, Villaescusa-Navarro2022, Ni+23}. In this paper, we also use a new suite of simulations, where the Universe volume is increased by eight times to $(50\,h^{-1}\;\textrm{Mpc})^{3}$. The specific details of the simulations, other than the volume, are already described in detail in Paper I and II. Each simulation has the following cosmological parameter values fixed: $\Omega_{\textrm{b}} = 0.049$, $n_{\textrm{s}} = 0.9624$ and $h = 0.6711$. We will limit to the \camels\ simulations based on the IllustrisTNG subgrid physics. The masses and radii of the galaxies investigated in this paper significantly exceed particle mass resolutions and softening length, respectively.

The fiducial \camels\ simulation uses the same parameters of IllustrisTNG, which is calibrated to observations. These calibrations have been performed by using the galaxy stellar mass function, the stellar-to-halo mass relation, the total gas mass content within the virial radius $r_{500}$ of massive groups, the stellar size - stellar mass and the black hole (BH) mass - galaxy mass relations, all at $z=0$, and, finally, the functional shape of the cosmic star formation rate density for $z\lesssim 10$ \citep{Pillepich2018}.

With respect to the original \camels\ suites, which allowed for the variation of only six parameters, the updated set of simulations allows us to explore the role of 28 astrophysical and cosmological parameters in total. In particular, we vary the two cosmological parameters ($\Omega_{\textrm{m}}$, $\sigma_{8}$), the originally implemented SN- and AGN-feedback parameters ($A_{\textrm{SN1}}$, $A_{\textrm{SN2}}$, $A_{\textrm{AGN1}}$, and $A_{\textrm{AGN2}}$) and six more parameters regulating AGN feedback. In particular, $A_{\textrm{SN1}}$ and $A_{\textrm{SN2}}$ are related to the SN feedback mechanisms, wind energy and velocity, respectively. Instead, $A_{\textrm{AGN1}}$ and $A_{\textrm{AGN2}}$ are the normalization factors of energy and frequency of AGN feedback \citep{Villaescusa-Navarro2021}. However, previous papers have demonstrated that $A_{\textrm{AGN1}}$ and $A_{\textrm{AGN2}}$ have no effect on scaling relations (see for example Papers I and II), and therefore we only investigate the impact of the six other AGN feedback parameters, which have been implemented in the latest \camels\ release. The complete list of parameters we will consider in this paper are listed in \Tab\ref{tab:parameters}. We use the so called '1P' simulations, where we can vary one of the parameters at a time, holding the rest of the parameters constant \citep[e.g.,][]{Villaescusa-Navarro2021}. The values of the simulation parameters varied in each '1P' simulation set (including the fiducial simulation ones) are listed in \Tab\ref{tab:params_values}.  In the rest of the paper, astrophysical parameter values are normalized to the parameter value of the fiducial simulation, which corresponds to the unit value in our figures.

Differently from our previous papers, where we analyzed different scaling relations, comparing simulations with observations, in the present paper we will only use the simulations and mostly the scaling relation $M_{\textrm{DM},1/2}/M_{\star,1/2}$ vs \mst\, where $M_{\textrm{DM},1/2}$ and $M_{\star,1/2}$ are the 3D DM and stellar mass calculated within the 3D stellar half-mass radius, respectively, \mst\ is the total stellar mass\footnote{We limit the analysis to a \cite{Chabrier03} IMF, deferring the investigation of the IMF's impact on stellar mass and DM fraction to a future paper (see more on the IMF in \citealt{Tortora+09,TRN13_SPIDER_IMF}).}, and $M_{\rm tot}$ is the total mass. For completeness, and for connecting the central galaxy regions to the whole galaxy, we also investigate the correlation between the total $\mst/M_{\rm DM}$ and stellar mass. We explore how these scaling relations are varying as a function of redshift, using simulations at the following redshift slices: 0, 0.5, 1.0, 1.5 and 2.

Similarly to the previous work, we performed a filtering of the subhalos detected by \textsc{subfind}. We considered only subhalos that have $R_{\star,1/2}>\epsilon_{\textrm{min}}$, $N_{\star,1/2}>50$ and $f_{\textrm{DM}}(<R_{\star,1/2}) \equiv 1 - M_{\star,1/2}/M_{\textrm{tot},1/2} > 0$, where $\epsilon_{\textrm{min}}=2\; \textrm{ckpc}$ is the gravitational softening length of the IllustrisTNG suite\footnote{In physical units, the softening length takes the following values: 2.0, 1.3, 0.978, 0.806, and 0.666 kpc at redshifts $z = 0$, $0.5$, $1.0$, $1.5$, and $2$, respectively.}. The selection criteria on $R_{\star,1/2}$ and $N_{\star,1/2}$ prevent strong resolution biases.

Similarly to Paper I and II, we follow \cite{Bisigello2020} for the selection of passive (PGs) and star-forming (SFGs) galaxies via specific SFR ($\textrm{sSFR}:=\textrm{SFR}/M_{\star}$), choosing for the PGs only subhalos having $\log_{10}(\textrm{sSFR}/\textrm{yr}^{-1})<-10.5$, and the opposite for SFGs. We have verified that varying the $\textrm{sSFR}$ threshold by $\pm 0.5\,\textrm{dex}$ does not affect our results (see Paper I)\footnote{Note that in our previous papers we adopted a different nomenclature—referring to early-type and late-type galaxies instead of passive and star-forming—even though the same selection criterion was applied. This choice was motivated by the direct comparison with observational data, which were classified primarily based on morphology rather than star formation activity.}. Although not a central focus of this paper, we also investigate the SFG main sequence and compare it with results from the literature in \App\ref{app:SFRmass_literature}.

The key observables from the simulations used in this paper are listed in \Tab\ref{tab:observables}.

\begin{table}
\centering
\caption{Observables used in the paper.}\label{tab:observables}
\begin{tabular}{cl}
\hline
\rm Observable & \rm Description\\
\hline
$R_{\star,1/2}$ & half-mass radius\\
\mst\ & stellar mass\\
$M_{\rm tot}$ & total mass\\
$M_{\textrm{tot},1/2}$ & total mass within $R_{\star,1/2}$\\
$M_{\textrm{DM},1/2}$ & DM mass within $R_{\star,1/2}$\\
$M_{\star,1/2}$ & stellar mass within $R_{\star,1/2}$\\
$N_{\star,1/2}$ & number of particles within $R_{\star,1/2}$\\
SFR & current star formation rate\\
\hline
\end{tabular}
\end{table}

\section{The golden mass}\label{sec:golden_mass}

Using the same parameters of the reference IllustrisTNG, calibrated, among other factors, to the observed $z=0$ size-stellar mass and halo-to-stellar mass relations, the fiducial \camels\ simulation reveals a bending in the simulated size-stellar mass and halo-to-stellar mass relations around a mass of approximately $5 \times 10^{10} \, \Msun$ (see also Paper I and II). This mass scale is transferred, for example, into the correlation among $M_{\textrm{DM},1/2}/M_{\star,1/2}$ and \mst\ and any other related scaling relation. Therefore this feature and the specific shape of these observed scaling relations are naturally and by construction introduced during the calibration of the reference simulation. However, since this calibration is not transferred to the other simulations with varied parameters, the scaling relations are no longer primarily influenced by the initial calibration. While the calibration process prevents a fully agnostic analysis of the emergence of the golden mass scale, it firmly enables us to assess the impact of cosmology and astrophysical processes on the shape of the scaling relations, as well as the emergence and value of the golden mass over cosmic time.

\begin{figure*}
\centering
\includegraphics[width=\linewidth]{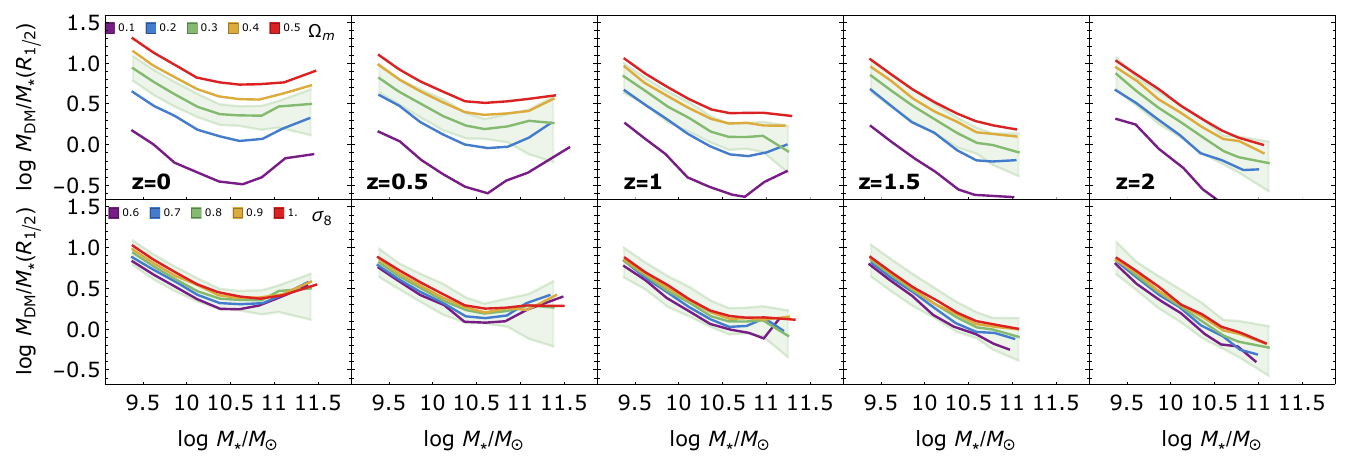}
\caption{The ratio $M_{\textrm{DM},1/2}/M_{\star,1/2}$ is plotted against \mst\ for \camels\ simulations, with each row showing variations in the cosmological parameters $\Omega_{\rm m}$ and $\sigma_{8}$. The green lines represent the reference \camels\ simulation. To examine the evolution across redshift, simulation snapshots at $z=0$, $z=0.5$, $z=1$, $z=1.5$ and $z=2$ are displayed from left to right. Medians in stellar mass bins are shown. For the reference simulation the 16-84th percentile range is also shown as a shaded green region.}
\label{fig:MtMstar_vs_Mst_Cosmparams}
\end{figure*}

\begin{figure*}
\centering
\includegraphics[width=\linewidth]{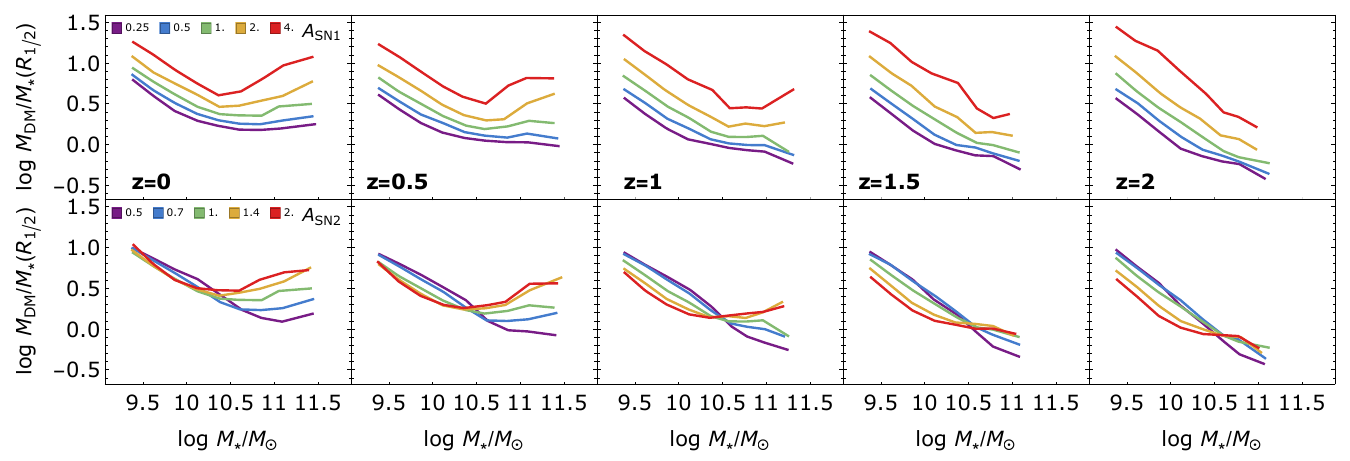}
\caption{The ratio $M_{\textrm{DM},1/2}/M_{\star,1/2}$ is plotted against \mst\ for \camels\ simulations, with each row showing variations in the SN-related parameters $A_{\rm SN1}$ and $A_{\rm SN2}$. The green lines represent the reference \camels\ simulation. To examine the evolution across redshift, simulation snapshots at $z=0$, $z=0.5$, $z=1$, $z=1.5$ and $z=2$ are displayed from left to right. Medians in stellar mass bins are shown.}
\label{fig:MtMstar_vs_Mst_ASNparams}
\end{figure*}

One significant limitation of the \camels\ simulations is the relatively small cosmological volume, which restricts the number of massive objects, especially in the mass range where the golden mass is expected to appear. However, compared to the original \camels\ simulations, which simulate a volume of $(25\,h^{-1}\,\textrm{Mpc})^{3}$, the current suite increases the volume, resulting in a galaxy count eight times larger. This increase in sample size is particularly important in the high-mass range ($\mst > 5 \times 10^{10} \, \rm \Msun$), where galaxies are less abundant, as it facilitates the determination of the golden mass.

To estimate the golden mass as a function of the simulation parameters, for each simulation we fit a parabola to the simulation data in the vicinity of the minimum in the scaling relation. The results for the $z=0$ snapshot are summarized in \Tab\ref{tab:golden_mass_values}\footnote{Note that using $M_{\rm tot,1/2}/M_{\star,1/2}$ instead of $M_{\rm DM,1/2}/M_{\star,1/2}$ increases the golden mass by less than 0.05–0.1 dex, without affecting our conclusions.}. The typical uncertainty in the (log) golden mass is approximately $0.1-0.2$ dex.

\subsection{Golden stellar mass}

\begin{figure*}
\centering
\includegraphics[width=0.95\linewidth]{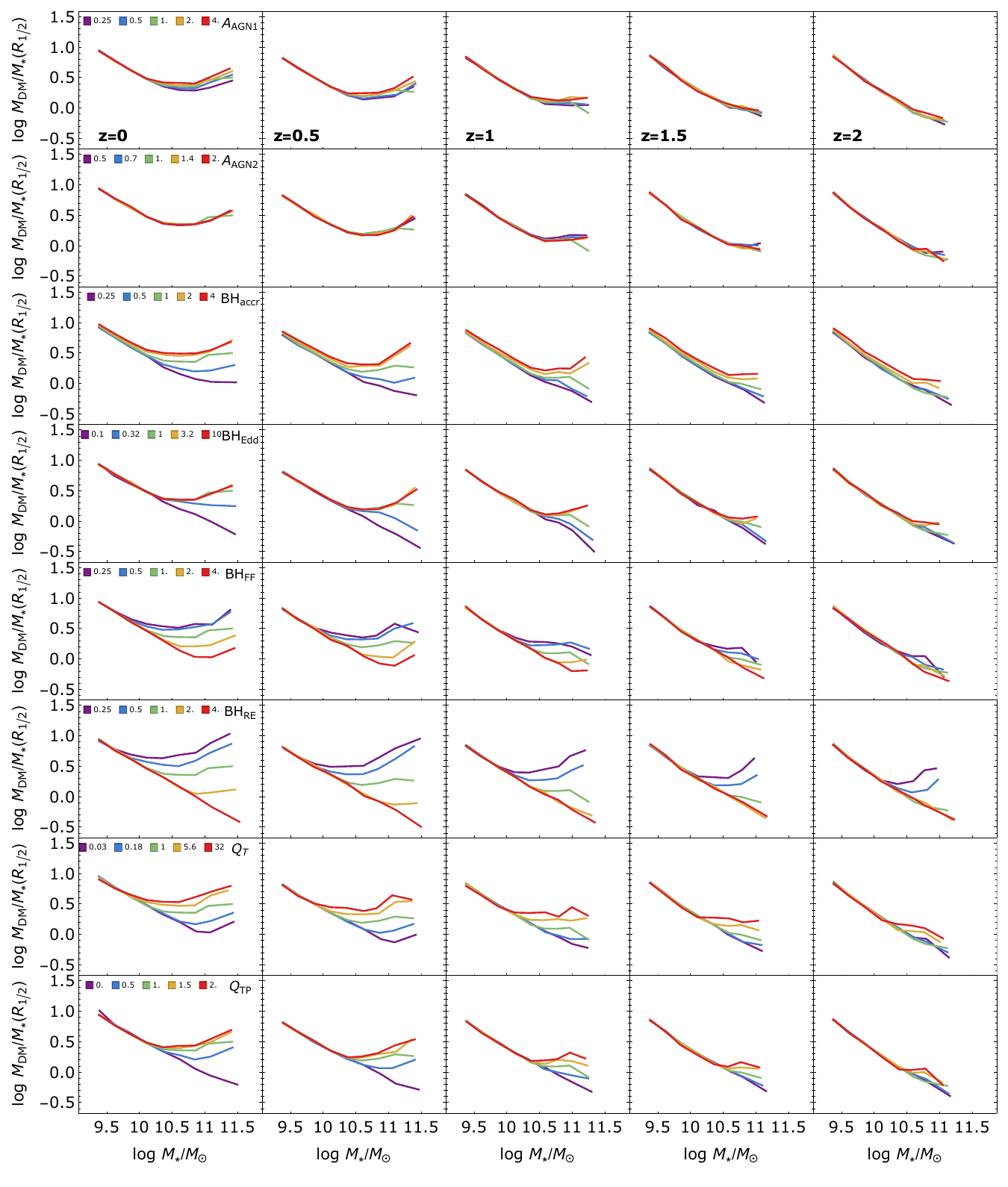}
\caption{The ratio $M_{\rm DM,1/2}/M_{\star,1/2}$ is plotted against \mst, with each row showing variations in the AGN-related simulation parameters: $A_{\textrm{AGN1}}$, $A_{\textrm{AGN2}}$, $BH_{\textrm{accr}}$, $BH_{\textrm{Edd}}$, $BH_{\textrm{FF}}$, $BH_{\textrm{RE}}$, $Q_{\textrm{T}}$, and $Q_{\textrm{TP}}$ from top to bottom. The green lines represent the reference \camels\ simulation. To examine the evolution across redshift, simulation snapshots at $z=0$, $z=0.5$, $z=1$, $z=1.5$ and $z=2$ are displayed from left to right. Medians in stellar mass bins are shown.}
\label{fig:MtMstar_vs_Mst_AGNparams}
\end{figure*}

We investigate the impact of the twelve simulation parameters listed in \Tab\ref{tab:parameters} on the $M_{\textrm{DM},1/2}/M_{\star,1/2}$--\mst\ scaling relation in \Figs\ref{fig:MtMstar_vs_Mst_Cosmparams}, \ref{fig:MtMstar_vs_Mst_ASNparams} and \ref{fig:MtMstar_vs_Mst_AGNparams}. Following the results presented in Paper II, for most of the simulations a U-shaped curve, with a minimum at the golden stellar mass, is observed. In particular, for the reference simulation, it takes the value of $\sim 5 \times 10^{10}\, \rm \Msun$. This result confirms what has been found combining observations for dwarf and giant early-types and late-types (\citealt{TLBN16_IMF_dwarfs,Tortora+19_LTGs_DM_and_slopes}) and not surprisingly results from Illustris TNG (\citealt{Lovell+18_Illustris}).

\begin{table}
\centering
\begin{threeparttable}
\caption{Golden mass estimates in $\log \mst/M_{\rm \odot}$, at $z=0$. The uncertainty on the estimated values is $\sim 0.1-0.2$ dex.}\label{tab:golden_mass_values}
%\resizebox{\textwidth}{!}{%
\begin{tabular}{ccccccc}
\hline
\rm Parameter & \multicolumn{5}{c}{Variations} & Trend \\
\rm & n2 & n1 & fiducial & p1 & p2 &  \\ 
\hline
$\Omega_{\rm m}$ & 10.5 & 10.6 & 10.6 & 10.7 & 10.7 & $\approx$ \\
$\sigma_{8}$ & 10.6 & 10.6 & 10.6 & 10.7 & 10.8  & $\approx$ \\
\hline
$A_{\rm SN1}$ &  10.8 & 10.7 & 10.6 & 10.6 & 10.5  & $\downarrow$ \\
$A_{\rm SN2}$ & 11.1 & 10.8 & 10.6 & 10.7 & 10.4  & $\downarrow \downarrow$ \\
\hline
$A_{\rm AGN1}$ & 10.8 & 10.7 & 10.6 & 10.6 & 10.6  & $\approx$ \\
$A_{\rm AGN2}$ & 10.6 & 10.7 & 10.6 & 10.7 & 10.7  & $\approx$ \\
$BH_{\rm accr}$ & - & 10.9 & 10.6 & 10.6 & 10.6  & $\downarrow$ \\
$BH_{\rm Edd}$ & - & - & 10.6 & 10.6 & 10.6  & $\approx$ \\
$BH_{\rm FF}$ & 10.5 & 10.5 & 10.6 & 10.8 & 11  & $\uparrow \uparrow$ \\
$BH_{\rm RE}$ & 10.3 & 10.5 & 10.6 & 11.1 & -  & $\uparrow \uparrow \uparrow$ \\
$Q_{\rm T}$ & 11 & 10.9 & 10.6 & 10.6 & 10.4 & $\downarrow \downarrow$ \\
$Q_{\rm TP}$ & - & 10.8 & 10.6 & 10.5 & 10.5  & $\downarrow$ \\
\hline
\end{tabular}
    \begin{tablenotes}
        \footnotesize
        \item $^1$ The parameter values corresponding to n2, n1, fiducial, p1, and p2, are listed in \Tab\ref{tab:params_values}.
        \item $^2$ The direction of the trend is shown in the last column, using the following symbols: $\approx$ indicates no statistically significant trend, $\downarrow$ and $\uparrow$ represent negative and positive correlations, respectively, while double or triple arrows denote stronger correlations.
    \end{tablenotes}
\end{threeparttable}
%}
\end{table}

\subsubsection{Cosmological parameters}

In \Fig\ref{fig:MtMstar_vs_Mst_Cosmparams}, where only the cosmological parameters are varied, and the rest of parameters are fixed to the values of the fiducial simulations, we show that the impact of $\Omega_{\rm m}$ on the scaling relation is more significant than that of $\sigma_{8}$ (reconfirming the findings reported in Papers I and II). Higher values of both parameters result in galaxies with greater DM content, although their impact on the golden mass appears negligible. We observe only a very mild flattening at the high-mass end for the largest $\sigma_{8}$ values. In general, we observe that the minimum in the scaling relation begins to emerge between $z=1.5$ and $z=1$, becoming definitively established around $z=0.5$.

\subsubsection{SN feedback parameters}

In \Fig\ref{fig:MtMstar_vs_Mst_ASNparams} we show the impact of SN feedback, by limiting the analysis to the two parameters $A_{\rm SN1}$ and $A_{\rm SN2}$, which determine the amplitude of the SN wind energy per unit star formation rate and the wind velocity, respectively. These are among the most influential parameters.

Consistent with findings in Papers I and II, we confirm that the SN energy per unit SFR parameter ($A_{\rm SN1}$) is one of the most significant astrophysical parameters affecting our scaling relation (Paper I).
Specifically, stronger wind energy corresponds to higher values of $M_{\rm DM,1/2}/M_{\star,1/2}$ at a fixed stellar mass. We can now also quantify the effect on the golden mass scale and the steepness of the scaling relation at masses below and above this scale. Notably, for low wind energy values, the bending is mild, and the golden mass becomes more pronounced as this parameter increases, becoming particularly well-defined for very strong wind energy. While the slope of the correlation at masses below the golden mass remains almost constant, the correlation for massive galaxies — dominated by PGs — becomes increasingly steeper as a function of wind energy. In the simulation with the strongest wind energy ($A_{\rm SN1} = 4$), the minimum—and thus the golden mass—occurs at $z \sim 1.5$. In contrast, simulations with weaker wind energy show that the minimum occurs at lower redshifts, with most models only revealing it at $z \sim 0$. In particular, at $z=0$, we observe a decrease in the golden mass as $A_{\rm SN1}$ increases, dropping from $7 \times 10^{10}\, \rm \Msun$ to $3 \times 10^{10}\, \rm \Msun$.

\begin{figure*}
\centering
    \includegraphics[width=\linewidth]{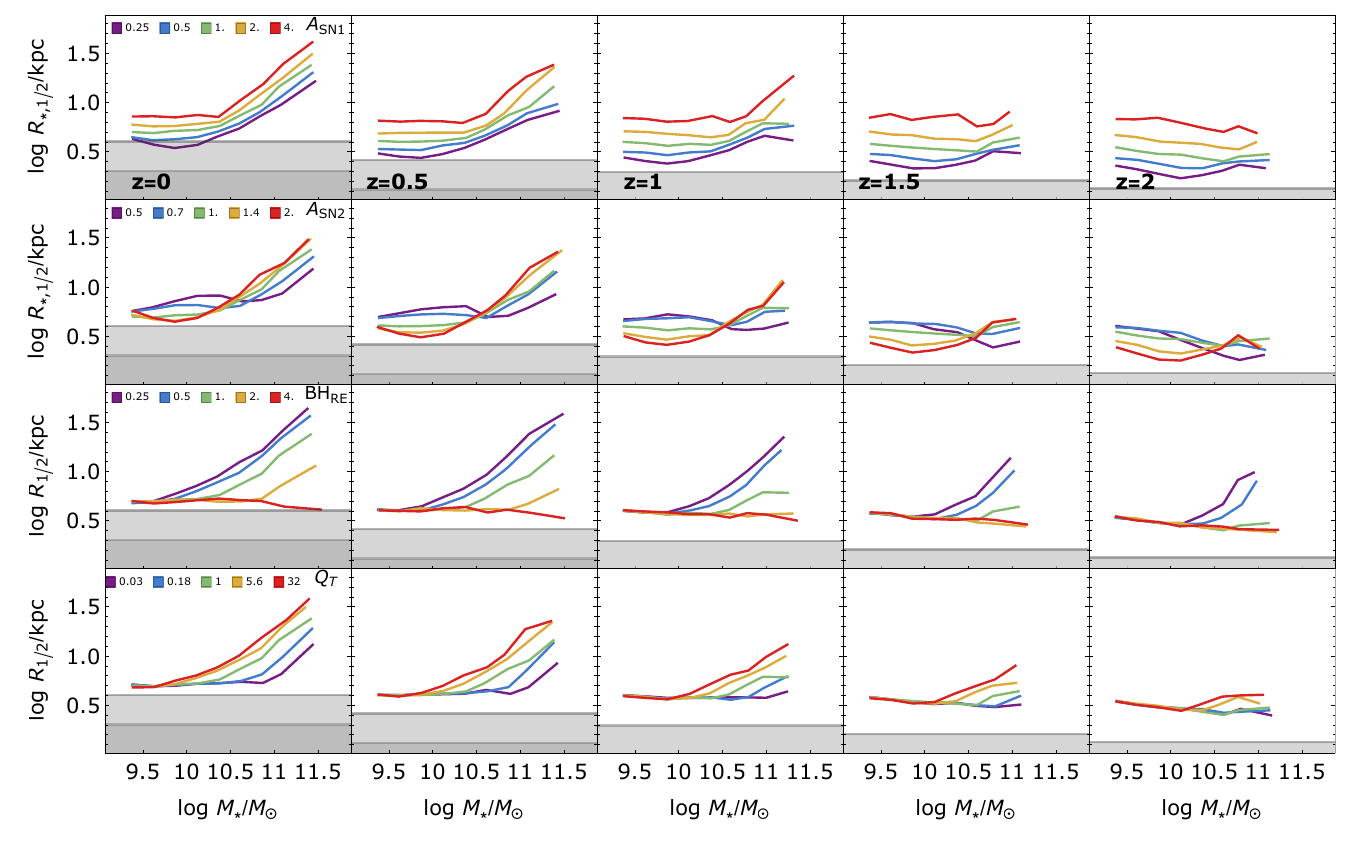}
\caption{The half-mass radius, $R_{*,1/2}$ is plotted against \mst, with rows showing variations in the two SN-related parameters $A_{\rm SN1}$ and $A_{\rm SN2}$ and two AGN parameters $BH_{\rm RE}$ and $Q_{\rm T}$. The green lines represent the reference \camels\ simulation. Simulation snapshots at $z=0$, $z=0.5$, $z=1$, $z=1.5$ and $z=2$ are displayed from left to right. Medians in stellar mass bins are shown. Dark and light gray regions indicate radii smaller than the softening length $\epsilon_{\rm min}$ and $2 \times \epsilon_{\rm min}$, respectively.}
\label{fig:Rh_vs_Mstar}
\end{figure*}

The impact of the wind velocity amplitude, $A_{\rm SN2}$, is more complex. At a fixed mass, as discussed in Paper II, the trend at high mass is similar to that of $A_{\rm SN1}$, with larger values corresponding to higher DM content. However, the trend is reversed at lower masses, where LTGs dominate. Regarding the emergence of the golden mass, similar to $A_{\rm SN1}$, the bending in the scaling relation is almost absent for the lowest wind velocity values, with the golden mass appearing at around $10^{11}\, \rm \Msun$. The bending becomes stronger with increasing wind velocity. A stronger decrease in the golden mass value is observed with increasing $A_{\rm SN2}$, ranging from approximately $10^{11}$ to $\sim 3 \times 10^{10}\, \rm \Msun$ for the extreme values of the parameter. When studying the correlations as a function of redshift, we find that a bend in the model with the highest wind velocity ($A_{\rm SN2} = 2$) is already present at $z = 2$, but the minimum in the relation only emerges at $z = 1$. In most models with lower $A_{\rm SN2}$ values, the golden mass appears at $z \lsim 0.5$.

\subsubsection{AGN feedback parameters}

The impact of eight parameters which regulate the AGN feedback in \camels\ simulations are finally investigated in \Fig\ref{fig:MtMstar_vs_Mst_AGNparams}. 

As noted in our previous papers (see also \citealt{Villaescusa-Navarro2022} and \citealt{Chawak2023}), the normalization factors for the energy and the frequency of events, i.e. $A_{\rm AGN1}$ and $A_{\rm AGN2}$, have a negligible effect on scaling relations, which we confirm for both the normalization of these relations and their impact on the golden mass scale. We only observe a minor effect on the slope of the trend in massive galaxies varying $A_{\rm AGN1}$ at $z < 0.5$.

In contrast, the impact of other AGN feedback parameters on the scaling relation is more significant, as it is expected to happen in the most massive galaxies, hosting super-massive BHs. 

Larger values of the BH accretion factor $BH_{\rm accr}$, BH Eddington factor $BH_{\rm Edd}$, quasar threshold $Q_{\rm T}$, and quasar threshold power $Q_{\rm TP}$ result in higher DM fractions at larger masses and introduce a bend in the scaling relation. For most of these parameters, and in simulations with their highest values, the golden mass is established around $z \sim 1$. For lower values of these parameters, the golden mass emerges at later redshifts. For BH accretion factor, quasar threshold, and quasar threshold power, larger golden mass values correspond to smaller parameter values. Focusing on the $z=0$ snapshot, smaller values of $BH_{\rm accr}$, i.e. smaller values for the normalization factor of the Bondi rate, push the golden mass to $\sim 10^{11}\, \rm M_\odot$, while it remains nearly stable for values greater than or equal to the fiducial. For $BH_{\rm Edd}$, models that show a clear minimum are those with values $\geq$ the fiducial, above which the golden mass stabilizes; therefore the normalization factor for the limiting Eddington rate is not impacting the golden mass value. $Q_{\rm T}$ regulates the threshold between the AGN feedback low- and high- accretion states,  and is one of the parameters with the greatest impact on the golden mass, reducing it from $10^{11}\, \rm M_\odot$ in the weaker models to $3 \times 10^{10}\, \rm M_\odot$ in the strongest models. Finally, the power-law index of the scaling relation of the low- to high-accretion state threshold with BH mass, i.e., the quasar threshold power ($Q_{\rm TP}$), shows a bending trend starting from $Q_{\rm TP} = 0.5$, with the golden mass and trends saturating for values $\geq$ the fiducial simulation.

The opposite trend is observed for the BH feedback factor $BH_{\rm FF}$, i.e. the normalization factor for the energy in AGN feedback in the high-accretion state, and the BH radiative efficiency $BH_{\rm RE}$, corresponding to the fraction of accretion rest mass released in the accretion process. In these cases, larger parameter values correspond to smaller DM fractions compared to lower values. For the smallest values of $BH_{\rm RE}$, the golden mass is already established at $z \sim 2$, while for larger values, it never appears. Regarding the BH feedback factor, the golden mass emerges across all models at $z < 1$. For both parameters, larger golden mass values correspond to larger parameter values. At $z=0$, although the BH feedback factor $BH_{\rm FF}$ has a greater impact on DM normalization, the golden mass saturates at the lower end of the parameter range, increasing beyond the fiducial model up to $\sim 10^{11}\, \rm \Msun$ for the highest parameter values. Finally, the fraction of accretion rest mass that is released in the accretion process proves to be the most influential parameter. Indeed, for the weakest values of $BH_{\rm RE}$, the golden mass is approximately $2 \times 10^{10}\, \rm M_\odot$, whereas for the largest values, it exceeds $10^{11}\, \rm M_\odot$.

\subsubsection{The connection with the size-mass relation}

We have seen that the golden mass is well defined and appears in earlier cosmic times in simulations with a stronger AGN and SN feedback, depending on the specific value of the parameters. These effect can also explained by a more rapid (or slower) size-mass evolution observed in these simulations.

By definition, the ratio  $M_{\rm DM,1/2}/M_{\star,1/2}$ depends on $R_{1/2}$, with larger DM fractions corresponding to larger $R_{1/2}$ values (e.g., \citealt{NRT10}). We therefore, as an example, present this correlation in terms of $A_{\rm SN1}$, $A_{\rm SN2}$, $BH_{\rm RE}$ and $Q_{\rm T}$ in \Fig\ref{fig:Rh_vs_Mstar}. The golden mass also emerges within these trends, although not as a minimum in the scaling relation, but as scale above which the scaling relation turns upward. At masses below the golden mass, the stellar half-mass radius remains nearly constant, while at larger masses, it becomes positively correlated with mass (except for the largest values of $BH_{\rm RE}$), consistent with the trends observed in the data (see Papers I and II; e.g., \citealt{Roy+18} and references therein). The trends with redshift indicate that galaxy sizes increase over cosmic time at all masses, but the effect is more pronounced at higher masses, where PGs dominate. For massive galaxies, the increase of the size (and consequently of the central DM fraction) is faster for models with stronger feedback (higher wind energy and velocity, smaller $BH_{\rm RE}$ and higher $Q_{\rm T}$), being the local size-mass relation already settled at earlier redshifts. The trends observed in relation to the SN-feedback parameters mirror those found for the central DM fraction in \Fig\ref{fig:MtMstar_vs_Mst_ASNparams}, where larger wind energies correspond to larger galaxy sizes. The characteristic bending associated with the golden mass is challenging to pinpoint, but it appears to be relatively unaffected. Since  $M_{\rm DM,1/2}/M_{\star,1/2}$ is positively correlated with $R_{1/2}$, the complex behavior noted in \Fig\ref{fig:MtMstar_vs_Mst_ASNparams} concerning $A_{\rm SN2}$ is also evident here: larger wind velocities lead to larger sizes at high masses and smaller sizes at lower masses. However, the impact of wind velocity on the golden mass seems consistent with that seen in DM fraction, with the bending occurring at smaller masses for higher wind velocities.

Interestingly, for both the correlations the golden mass is absent at $z=2$ for any set of parameters, except for extreme values of the AGN-feedback parameters. It first emerges at $z \sim 1-1.5$ for high values of wind energy and velocity, and only at $z=0$ for the lowest values of SN parameters. For extreme values of some AGN feedback parameters it emerges at $z \sim 2$, for other at lower redshifts, in some cases it never appears. As an example, it emerges at $z=1.5-2$ only for the smallest values of $BH_{\rm RE}$, and only at $z=0$ for low values of $BH_{\rm accr}$ or $Q_{\rm T}$ and high values of $BH_{\rm RE}$. This is suggesting a link to the quenching of galaxies. In all cases where the golden mass is observed, it becomes more consolidated over cosmic time. This emergence is also related to the behavior of the most massive galaxies, which begin to show a positive correlation between DM fraction and mass only between $z=1$ and $z=2$. Stronger feedback halts star formation, creating galaxies with smaller stellar masses and leading to a higher DM fraction. Star formation quenching and mergers contribute to the increase in size and DM fraction in the most massive galaxies over cosmic time. A more systematic analysis of the size-mass relation as a function of redshift, and the comparison with observations, are beyond the scope of this paper and will be analyzed in future papers.

Although we have selected only galaxies with $R_{\star,1/2} > \epsilon_{\textrm{min}}$, some residual resolution effects may still persist above this threshold. To assess their potential impact, in \Fig\ref{fig:Rh_vs_Mstar} we highlight  sizes below the softening length (excluded from our sample) and those below twice the softening length, across different redshifts. Overall, the vast majority of simulations—across stellar masses and redshifts—would not be affected by resolution limitations. However, exceptions arise in specific cases, such as simulations with very low values of $A_{\rm SN1}$ or very high values of $A_{\rm SN2}$ at $z \lesssim 0.5$. Overall, these exceptions may impact the scaling relations at the low-mass end—below the golden mass—but they do not affect our main conclusions.

\begin{figure*}
\centering
    \includegraphics[width=\linewidth]{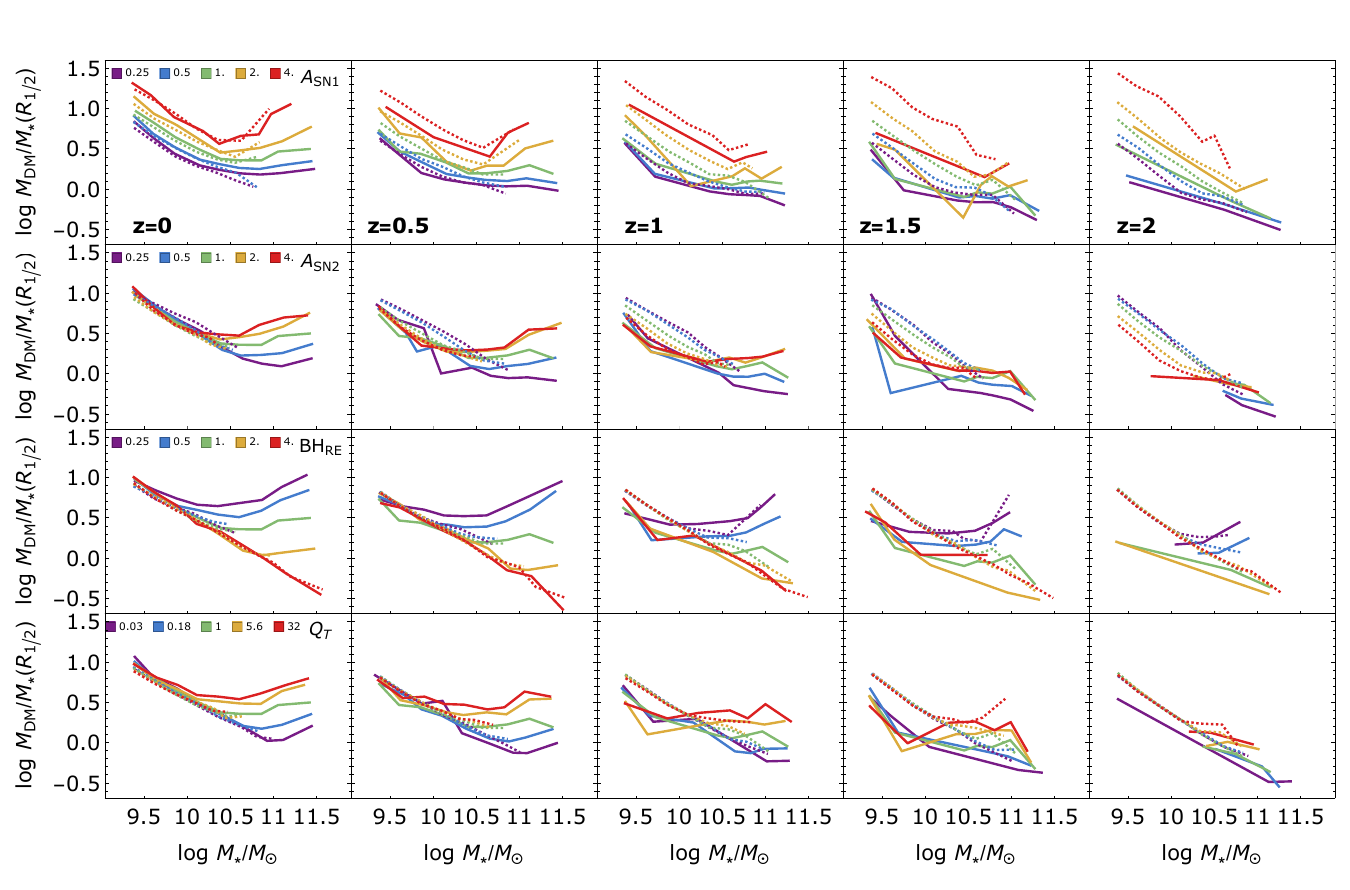}
\caption{The ratio $M_{\rm DM,1/2}/M_{\star,1/2}$ is plotted against \mst, with rows showing variations in the two SN-related parameters $A_{\rm SN1}$ and $A_{\rm SN2}$ and the two AGN parameters $B_{\rm RE}$ and $Q_{\rm T}$. The green lines represent the reference \camels\ simulation. Simulation snapshots at $z=0$, $z=0.5$, $z=1$, $z=1.5$ and $z=2$ are displayed from left to right. The sample is divided in PGs (solid lines) and SFGs (dashed lines). Medians in stellar mass bins are shown. The binning in the two subclasses has been optimized based on the number of points in the bin, and trends based on very few data points have been omitted.}
\label{fig:MtMstar_vs_Mstar_types}
\end{figure*}

\begin{figure*}
\centering
    \includegraphics[width=\linewidth]{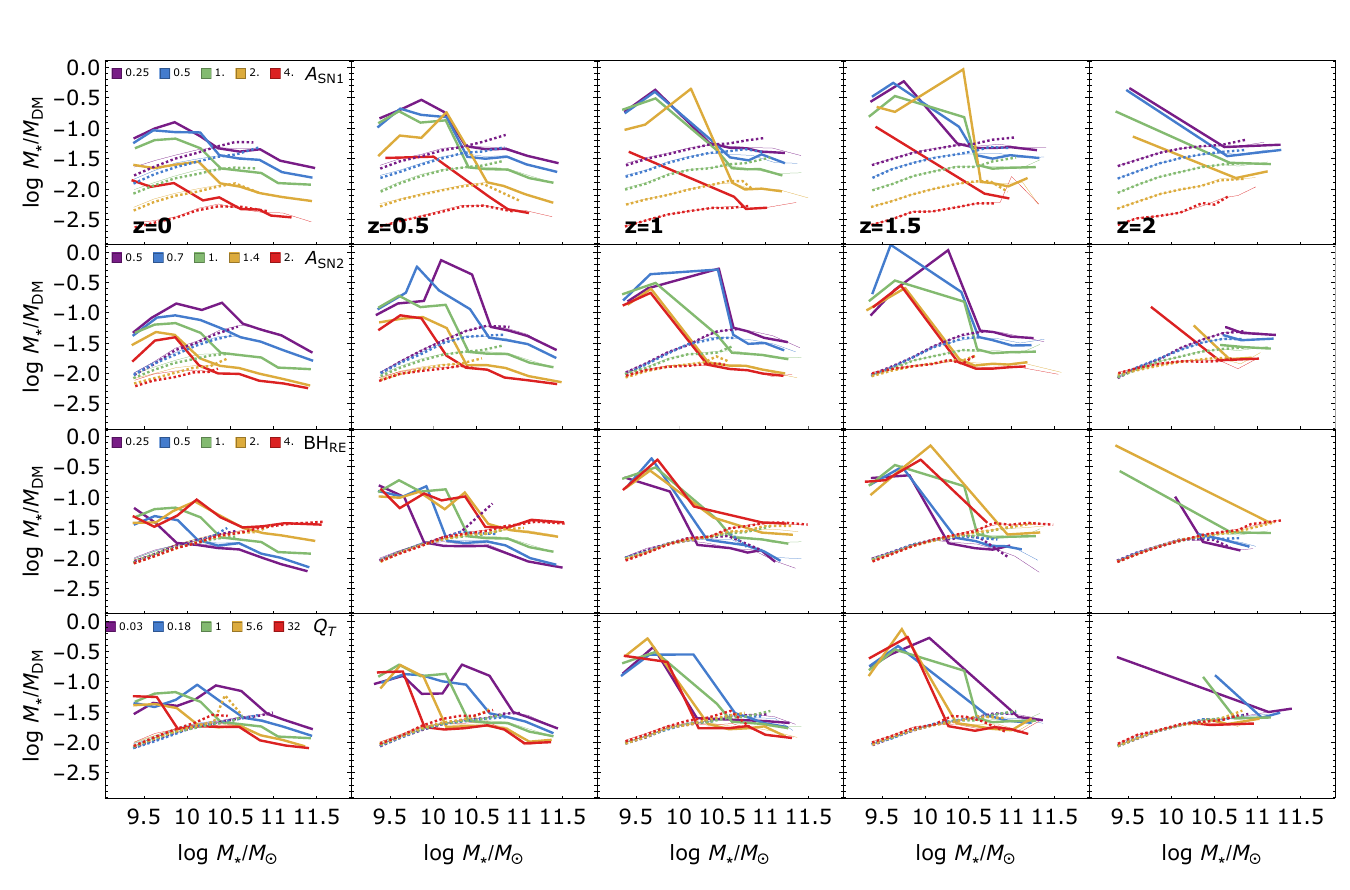}
\caption{The ratio $M_{\star}/M_{\textrm{DM}}$ is plotted against \mst, with rows showing variations in the two SN-related parameters $A_{\rm SN1}$ and $A_{\rm SN2}$ and two AGN parameters $BH_{\rm RE}$ and $Q_{\rm T}$. The green lines represent the reference \camels\ simulation. Simulation snapshots at $z=0$, $z=0.5$, $z=1$, $z=1.5$ and $z=2$ are displayed from left to right. The sample is divided in PGs (solid lines) and SFGs (dashed lines). Thin lines are used for the median of the full sample. Medians in stellar mass bins are shown.}
\label{fig:fstar_vs_Mstar_types}
\end{figure*}

\subsection{Galaxy type dependency}

Different types of galaxies are influenced by different processes and span different ranges of masses. Therefore, it is crucial to investigate whether scaling relations differ between these galaxy types. Figure \ref{fig:MtMstar_vs_Mstar_types} shows the relation between $M_{\rm DM,1/2}/M_{\star,1/2}$ and \mst\ for PGs and SFGs.

It is not surprising that the scaling relation for masses larger than the golden mass is predominantly driven by PGs. This is because the number density of star-forming galaxies decreases significantly at $\sim 10^{10.5-11}\, \rm \Msun$. Additionally, due to our selection criterion of constant sSFR with redshift, the number of PGs at $z=2$ is notably low.

For any values of the parameters shown, the variation in central DM fraction with redshift is weaker for SFGs compared to PGs. The stronger variation observed in PGs is favoured by their greater size evolution. Except for the $z=0$ snapshot, at fixed stellar mass, the DM fractions of SFGs are generally higher than those of PGs. The impact of SN feedback parameters remains consistent when galaxy types are explicitly considered. The trends observed for the entire sample at high masses apply to PGs, while some differences emerge at low masses and for SFGs.

In SFGs, the decreasing trend with \mst\ of the DM fraction is confirming the results obtained with data both at low- and high-z \citep{Tortora+19_LTGs_DM_and_slopes,Sharma2022}. Although the number density of SFGs approaches zero at high masses, a slight change in slope and the emergence of the golden mass is observed for extreme values of some parameters\footnote{In some panels, the same binning used across simulations obscures the detection of this bending.}. Some mild evidence of bending was found in \cite{Tortora+19_LTGs_DM_and_slopes}, using SPARC star forming galaxies with measured rotation curves (\citealt{Lelli+16_SPARC}), tending to exclude simulations with very extreme values of SN- and AGN-feedback parameters. However, the presence of the bend found in observations contrasts with findings in \cite{Posti2019}, which, using the sample used in \cite{Tortora+19_LTGs_DM_and_slopes}, reported (relying on global quantities) no bending in the relation between stellar fraction (of star formation efficiency) and stellar mass in the most massive spirals.

To resolve this discrepancy or at least discover the possible cause, it is important to examine the simulations' predictions for the correlation between the total $M_{\star}/M_{\textrm{DM}}$ and $M_{\star}$, the stellar-to-halo mass relation, obtained including all particles in the subhalos. The results are presented in \Fig\ref{fig:fstar_vs_Mstar_types}. 

We confirm the maximum in the relation when all galaxies are considered (e.g., \citealt{Moster+10}), which mirrors the trends observed for central DM fraction as a function of stellar mass (e.g., \Fig\ref{fig:MtMstar_vs_Mst_ASNparams}). The peak of this relation, or the golden mass, increases with higher redshifts. In the highest redshift bin, we observe a flattening at high masses. Models with higher wind energy and velocity tend to have larger central DM fraction, and lower global stellar fractions $\mst/M_{\rm DM}$, indicating lower star formation efficiency. Similarly to \Fig\ref{fig:MtMstar_vs_Mst_ASNparams}, varying $A_{\rm SN2}$ influences the golden mass, whereas $A_{\rm SN1}$ less. Similitude with \Fig\ref{fig:MtMstar_vs_Mst_AGNparams} can also be made by comparing the trends for AGN-feedback parameters shown in terms of scaling relation normalization and variation in the golden mass value. Limitations due to resolution and simulation volume are investigated in \App\ref{app:SGMR_literature}, where we also compare with some semi-empirical SHMR from the literature. Except for some shift induced by the smaller resolution of our simulations and the lower number of massive galaxies, the results are quite conform to the original TNG simulations, while some disagreement emerge between both \camels\ and TNG simulations with literature results using halo abundance matching \citep{Moster+10,Moster2013,Girelli+20}.

For SFGs alone, indications of a golden mass are observed (although very weak), for example in models with high $A_{\rm SN1}$ and $Q_{\rm T}$ values, which contradicts the findings of \cite{Posti2019}. The absence of bending in the \cite{Posti+19} results would favour models with a weaker feedback. However, a larger volume would be beneficial to better sample the most massive end of the mass function where LTGs are extremely rare.

An interesting result pertains to the PG population. When considering global quantities, the stellar fraction in our case, no clear bending at the golden mass is observed. In contrast, there are mild indications of a peak at lower masses ($\sim 10^{9.5-10}\, \rm \Msun$), which increases for larger (smaller) values of, for example, $BH_{\rm RE}$ ($Q_{\rm T}$). This contrasts with the analysis of the central quantities, where the golden mass clearly emerges for the whole sample and PGs alone, driven by the bending observed in the size-mass relation in both samples (\Fig\ref{fig:MtMstar_vs_Mstar_types}).  This may be attributed to the fact that, on global scales, SN and AGN feedback have not yet had sufficient time to take effect.

At low stellar masses ($\lsim 10^{10-10.5}\, \rm \Msun$) PGs are characterized by lower halo masses with respect to SFGs of similar stellar masses. Instead, at larger masses halo masses of the two types are quite similar, with many simulations showing some mild indications of larger halo masses in SFGs. Except for the highest-masses in some of our simulations, for the rest we find results opposite to the literature \citep{Rodriguez-Puebla+15,Correa_Schaye2020}. These results are qualitatively unaffected by a change in the sSFR threshold. We have also explored the impact on the trends of a different classification based on the ratio $V_{\rm max}/\sigma$, finding that, both cold late-type (low $V_{\rm max}/\sigma$) and hot early-type (high $V_{\rm max}/\sigma$) systems follow a similar trend with stellar mass, with larger stellar fractions in the former, which looks more consistent with results from \cite{Rodriguez-Puebla+15}. In future analyses, we will explore the origin of this result in greater detail.

\section{Discussion}\label{sec:discussion}

\begin{figure}
\centering
\includegraphics[width=\linewidth]{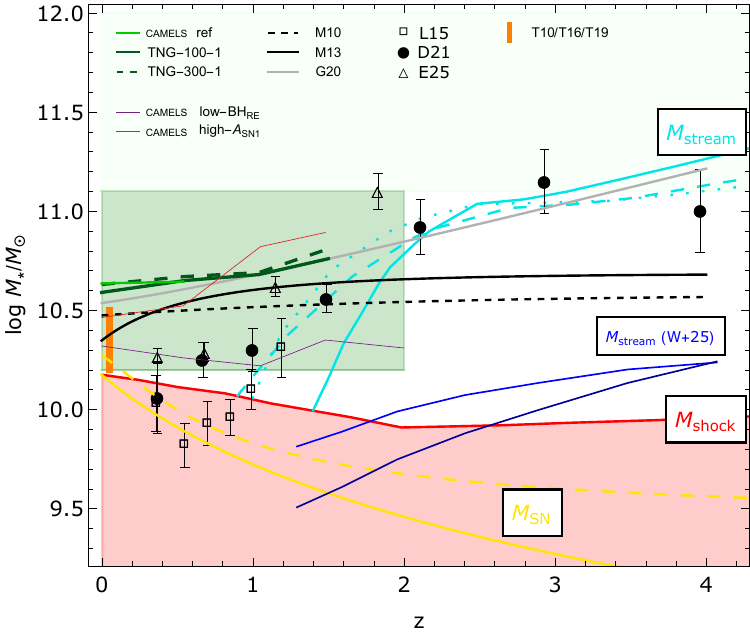}
\caption{Golden mass determined from the central $M_{\rm DM}/\mst - \mst$ relation assuming the reference \camels\ simulation (green line) is compared with other results from the literature. A qualitative range for the golden mass values determined from all the \camels\ simulations explored in this paper are shown as dark and light green regions, see text for more details. We also plot the results for the simulations corresponding to the lowest $BH_{\rm RE}$ and highest $A_{\rm SN1}$ values. Then, we compare with the golden mass estimated from the same correlation for TNG100-1 (dark green line) and TNG300-1 (dashed dark green line) simulations. From the theoretical point of view, we also show the shock heating stellar mass ($M_{\rm shock}$) as a red line and the stream mass ($M_{\rm stream}$) as solid cyan line \citep{dek_birn06}, together with its two revised forms (dashed and dotted cyan lines) discussed in \cite{Daddi+22_main_sequence}. These curves for $M_{\rm shock}$ and $M_{\rm stream}$ are extracted from \cite{Daddi+22_main_sequence}, where more details can be found. The stream mass determined in \cite{Waterval+25} is also plotted as blue and dark blue lines (see more details in the text). We show the typical mass above which SN feedback interrupts its quenching effect, using two conversions from halo to stellar mass, see text for more details ($M_{\rm SN}$, yellow lines). From the observational point of view, we plot the range of golden masses in the local Universe from different observables \citep{Tortora+10CG,TLBN16_IMF_dwarfs,Tortora+19_LTGs_DM_and_slopes}. Then, we also plot the stellar mass value of the bending in the SHMR determined from using halo-abundance matching in \citet[][black dashed line]{Moster+10}, \citet[][black solid line]{Moster2013} and \citet[][grey solid line]{Girelli+20}, and the bending mass in the main sequence of star-forming galaxies \citep{Lee+15,delvecchio+21,Enia+25_Euclid}.}
\label{fig:comparisons_literature}
\end{figure}

The \camels\ simulations are constructed by varying a range of parameters associated with SN and AGN feedback. We have leveraged this flexibility to investigate how these parameters influence the $M_{\rm DM}/\mst$–\mst\ scaling relation and the emergence of the golden mass, defined as the stellar mass corresponding to the minimum of this relation, across different redshifts.

Our analysis shows that cosmological parameters do not affect the golden mass. In contrast, several feedback-related parameters—particularly those governing supernova wind velocity and energy, as well as AGN feedback (in particular the fraction of accreted rest mass energy released, the Eddington ratio, and the normalization factor for the AGN energy feedback in the high-accretion state)—play a significant role in shaping this feature. Except in cases where the fraction of accretion rest mass ($BH_{\rm RE}$) is extremely low, the minimum in the scaling relation persists, allowing the golden mass to be identified up to $z \sim 2$. For the remaining parameter variations, the golden mass consistently emerges from $z = 1$ onward.

Before proceeding, we caution that beyond $z \sim 1$, the limited simulation volume leads to a sharp decline in the number of galaxies with stellar masses above $\sim 10^{11}\, \rm \Msun$. This scarcity makes the identification of the golden mass increasingly uncertain—and potentially infeasible—at higher redshifts. Addressing this limitation requires simulations with larger volumes. To explore this issue, we derived the corresponding scaling relations using the Illustris TNG100-1 and TNG-300-1 simulations \citep{Nelson2019a}, which contain .approximately 3.4 and 69 times more galaxies than the \camels\ simulations analyzed here, in \App\ref{app:SGMR_literature}. We find that the shape of the scaling relations and the position of the golden mass are broadly consistent between TNG and \camels\ simulations. A similarity between TNG300-1 and \camels\ is expected due to their comparable resolution; however, it is particularly encouraging that the golden mass remains consistent even in the higher-resolution TNG100-1 simulation. The larger volumes of TNG100-1 and TNG300-1, however, enables the golden mass to be probed more robustly at $z \gsim 1$. 

In this section we collect and present the results of the paper using \camels\ simulations in \Fig\ref{fig:comparisons_literature}, where our golden mass estimates are compared with results from other simulations and different observational data. We will discuss these results in the context of virial shocks, cold gas streams, and feedback from SN and AGN.

\subsection{Different mass scales from simulations and observations}

As discussed in \Sec\ref{sec:state-of-the-art}, simulations suggest that above a virial mass of $M_{\rm shock} \, \sim 10^{11.7}\, \rm \Msun$, virial shocks heat the CGM, thereby suppressing new star formation. However, at $z \gtrsim 1.5$, cold gas streams are expected to continue fueling star formation in halos below a characteristic mass scale, $M_{\rm stream}$, which increases with redshift. Consequently, we expect galaxies to be predominantly passive above $M_{\rm shock}$ at $z \lesssim 1.5$, and above $M_{\rm stream}$ at higher redshifts. These trends are illustrated in \Fig\ref{fig:comparisons_literature}, where halo masses $M_{\rm shock}$ and $M_{\rm stream}$ have been converted into stellar masses following \citet[][Figure 2, left panel]{Daddi+22_main_sequence}. For completeness, we also include the results from \cite{Waterval+25}, who use cosmological hydrodynamical simulations from the High-z Evolution of Large and Luminous Objects (HELLO) and the Numerical Investigation of a Hundred Astrophysical Objects (NIHAO) projects. We convert the stream halo mass trend in their Figure 14 into stellar mass, using the SHMR in \citet[][blue]{Girelli+20} and \citet[][darker blue]{Moster2013}. Interestingly the stream mass in \cite{Waterval+25} is about 1 dex smaller than the \cite{dek_birn06} and \cite{Daddi+22_main_sequence} relations. In the following we will mostly refer to the \cite{dek_birn06} and \cite{Daddi+22_main_sequence} relations.

By analyzing the bending of the SFR–stellar mass relation along the main sequence of star-forming galaxies, \citet{Daddi+22_main_sequence}—building on measurements from \citet{Lee+15} and \citet{delvecchio+21}—found that the characteristic bending mass increases with redshift, from $\sim 10^{10}\, \rm \Msun$ at $z \sim 0.35$ to $\sim 10^{11}\, \rm \Msun$ at $z \sim 2$ and beyond (see also \citealt{Enia+25_Euclid} for consistent results). Their bending mass seems to follow the redshift evolution of $M_{\rm stream}$ (or its revised versions obtained in \citealt{Daddi+22_main_sequence}). Their analysis suggests that the main sequence bending is driven by a decline in cold gas accretion, which limits the fuel available for SF—a scenario consistent with the cold-stream model. Above the $M_{\rm shock}/M_{\rm stream}$ mass threshold, the gas supply dwindles, leading to a plateau in the star formation rate rather than an abrupt quenching. Importantly, \citet{Daddi+22_main_sequence} argue that quenching is governed by additional physical mechanisms beyond the suppression of cold accretion. They also emphasize that the characteristic mass associated with the bending of the main sequence differs from the knee of the stellar mass function, which tends to be higher at $z \lesssim 1.5$.

However, their study does not include a direct measurement of the main sequence bending at $z \sim 0$. By extrapolating the observed trends, we may estimate that the bending mass at $z \sim 0$ likely falls within the range $\sim 10^{9.5}$–$10^{10}\, \rm \Msun$. Interestingly, this scale seems to be different from the typical golden mass values quoted in the literature (see \Sec\ref{sec:state-of-the-art}). The bending mass, instead, resembles the first mass scale proposed by \citet{Kannappan+13}, who identified a transition mass at $z \sim 0$ separating an "accretion-dominated" regime from a "processing-dominated" one, where the galaxies start processing the remaining available gas reservoir. These initial considerations prompt us to raise a cautionary note regarding the definition of mass scales, as the transition mass observed in the star-forming main sequence—identified as the "gas-richness threshold" by \citet{Kannappan+13}—may not correspond to the same physical scale as the transition mass investigated in this work, which resembles the "bimodality mass" described in the same study.

In fact, plotting the stellar mass corresponding to the bending of the SHMR results from halo abundance matching \citep{Moster+10,Moster2013,Girelli+20}, we see that this mass value aligns better with numerous observational findings in the local Universe, particularly with the range of golden mass values identified through minima in color gradients \citep{Tortora+10CG}, as well as through trends in stellar mass of total mass density slopes and DM fractions \citep{TLBN16_IMF_dwarfs, Tortora+19_LTGs_DM_and_slopes}. 

Looking at the evolution with redshift, contrasting results are found. We see a mild evolution in \cite{Moster+10} and \cite{Moster2013}, with a flattening at $z \gsim 1$, A notable discrepancy arises when comparing these values with the shock and stream transition masses $M_{\rm shock}$ and $M_{\rm stream}$. While at $z \lesssim 1.5$ the star formation efficiency peak lies above $M_{\rm shock}$, at higher redshifts it falls between $M_{\rm shock}$ and $M_{\rm stream}$—a regime where cold gas flows are still expected to sustain star formation. This contrasts with the consideration that the peak of the star formation efficiency separate mostly passive and star forming systems. Differently, the more recent results in \cite{Girelli+20}, present a stronger evolution with redshift, reaching a value of  $\sim 10^{11}\, \rm \Msun$ at $z \sim 3$, and aligning with the $M_{\rm stream}$ redshift evolution above $z \sim 2$. The SHMR for these papers are plotted in \Fig\ref{fig:fig_SHMR_literature}.

In this work, we have provided a new set of predictions for the golden mass, identified as the minimum in the $M_{\rm DM}/\mst$–\mst\ scaling relation. Across our suite of simulations, this minimum typically falls within the range $\sim 10^{10.3}$–$10^{11.1} \, \rm M_\odot$ at $z \lesssim 2$ (darker green region in \Fig\ref{fig:comparisons_literature}, \Tab\ref{tab:golden_mass_values}). We cannot exclude that in few extreme cases the mass can extend to even higher masses—though these lie in a regime where our simulations become increasingly limited due to low statistics (lighter green region in \Fig\ref{fig:comparisons_literature}). At higher redshifts, most simulations do not exhibit a clear minimum in the relation, aside from a few outliers. This absence may reflect the genuine disappearance of the golden mass at early times or may indicate that its detection requires larger-volume simulations capable of massive systems with $\mst \gtrsim 10^{11} \, \rm M_\odot$.

In particular, to investigate the impact of the low galaxy statistics at large mass, we compare the trends from the \camels\ reference run with the TNG300-1 and TNG100-1 simulations, obtained with the central $M_{\rm DM}/\mst$–\mst\ relation. The SHMRs for these simulations are compared in \Fig\ref{fig:fig_SHMR_literature}. In \Fig\ref{fig:comparisons_literature} we can see that, as expected, the golden mass for \camels, TNG100-1 and TNG300-1 are quite the same, confirming the goodness of \camels\ simulations, although the limited volume. Therefore, at $z \sim 0$ \camels\ and TNG simulations predict golden mass values that are systematically $0.1$–$0.2$ dex higher than those inferred from observations \citep{Tortora+10CG, Tortora+19_LTGs_DM_and_slopes, Moster2013}, and far more larger than the bending mass in the star formation main-sequence. The trends with reshifts look quite similar to halo abundance matching results in \cite{Girelli+20}. Also using the higher volume and resolution of TNG100-1 and the much more larger volume of TNG300-1, at $z \gsim 1.5-2$ no golden mass is found, therefore it is natural to expect the same in simulations with similar or weaker SN- and AGN-feedback, and no conclusion can be drawn for extreme strong feedback models.

Thanks to their larger volumes, the golden mass can be traced in TNG300-1 and TNG100-1 up to $z \sim 1.5$, compared to the more limited range of $z \sim 0.5-1$ accessible with the reference \camels\ simulation. These findings suggest that future studies would benefit from running \camels\ simulations in larger volumes, which would enable more robust tracking of redshift evolution and better characterization of the impact of varying SN and AGN feedback parameters.

The range of the golden mass in \camels\ and TNG simulations however seems to be within region above $M_{\rm shock}/M_{\rm stream}$, where the availability of cold gas is limited.

\subsection{Feedback, shocks and streams: competing or alternative processes?}

But do these results imply that our simulations are consistent with the shock/stream scenario? Are virial shocks and cold streams the fundamental drivers behind the emergence of the golden mass in the scaling relation? Or are these processes acting in a complementary way? Furthermore, could discrepancies arise from the use of different mass definitions across studies?

We have shown and discussed that SN and AGN feedback parameters have a significant impact on the emergence of the golden mass. In particular, when the strength of SN and AGN feedback is reduced, the minimum in the scaling relations—even at $z \sim 0$—tends to disappear. While we do not have access to simulations in which these feedback mechanisms are completely turned off, the trends observed in our results strongly suggest that such a scenario would likely prevent the formation of a well-defined golden mass. This reasoning is performed at $z \lsim 0.5$, where limitations due to the simulated volume are not affecting our inferences. However, these considerations seem to also hold at larger redshifts up to $z \sim 1.5-2$, considering the higher volume and resolution of the original TNG simulations. We can extrapolate these results to hypothetical TNG simulations with weaker feedback models at $z \gsim 1.5-2$. This indicates that, on their own, shock and stream processes alone may not be sufficient to produce the minima seen in our scaling relations—at least within the \camels\ framework using the Illustris TNG subgrid model.

Further analysis is needed to disentangle the roles of these two feedback processes. At low stellar masses, SN feedback is known to be the dominant mechanism regulating star formation \citep{Nelson+19_TNG, Weinberger+20_TNG_BHfeed}. Theoretical considerations \citep[e.g.,][]{Dekel_Silk86} provide an estimate for the upper halo mass where SN feedback remains effective, $M_{\rm SN} = M_{\rm SN,0} (1+z)^{-3/2}$. Assuming $M_{\rm SN,0} = 10^{11.7} \, \rm \Msun$, after converting this halo mass to a stellar mass, following \cite{Daddi+22_main_sequence}, we find a corresponding value of $\sim 10^{10.2}\, \rm M_\odot$ at $z = 0$, which decreases rapidly with redshift to $\sim 10^{9.5}\, \rm M_\odot$ at $z \sim 2$ (shown as solid yellow line in \Fig\ref{fig:comparisons_literature}). A slightly different trend is found if the \cite{Moster2013} SHMR is used to convert from halo to stellar mass (dashed yellow line). However, this trend disagrees with our findings, since in \Fig\ref{fig:MtMstar_vs_Mst_ASNparams}, we also see that modifying SN feedback parameters can influence the scaling relation across a broad mass range, up to $\mst \sim 10^{11.5}\, \rm M_\odot$. While the effect of wind velocity appears to diminish in massive galaxies at high redshift, wind energy continues to have a strong impact on galaxies at all masses and redshifts. This mass scale is much more larger than the $M_{\rm SN}$ mass estimated by the above-mentioned theoretical calculations.

AGN feedback, in contrast, becomes increasingly dominant at higher stellar masses. It not only expels gas from central regions but also heats it significantly, driving powerful outflows, as shown in TNG simulations \citep{Nelson+19_TNG, Weinberger+20_TNG_BHfeed, Terrazas+20_TNG_BH, Zinger+20_TNG}. Our parameter studies confirm that the influence of AGN feedback grows with mass, becoming particularly relevant at $\mst \gtrsim 10^{10}\, \rm \Msun$ in most of the models examined.

Finally, can we constrain the strength of feedback processes? At this stage, we cannot fully adopt the methodology used in the previous CASCO papers, as we do not yet have the large ensemble of \camels\ simulations in which cosmological and astrophysical parameters are finely, uniformly and randomly sampled. Instead, we are limited to the 1P simulations, where only one parameter is varied at a time. We defer a more robust statistical analysis to a future study, once the full set of simulations becomes available. Nonetheless, we can offer some qualitative constraints by comparing the golden mass derived from the 1P simulations with observational benchmarks. Currently, central DM-to-stellar mass relations are primarily available at $z \sim 0$, while at higher redshifts the limited mass range probed by observations makes it difficult to constrain the bending mass systematically. For this reason, we focus on the local Universe, comparing our results to the typical golden mass inferred from local central DM-to-stellar mass relations and other datasets (represented by the thick orange vertical line in \Fig\ref{fig:comparisons_literature}).

A comparison with the higher-resolution reference TNG100-1 simulation shows that resolution has only a mild effect on the inferred golden mass in \camels, which allows us to neglect this source of systematics, which seems to have a larger effect on the normalization of the SHMR at fixed stellar mass, as seen in \Fig\ref{fig:fig_SHMR_literature}. To match the observed golden mass values at $z \sim 0$ or approach them, one would need to strengthen the effects of SN or AGN feedbacks. This could be achieved, for instance, by maximizing one of the SN feedback parameters ($A_{\rm SN1}$ or $A_{\rm SN2}$), reducing the value of the BH radiative efficiency $BH_{\rm RE}$ to its lowest simulated level, or increasing the quasar threshold $Q_{\rm T}$. The results for the highest $A_{\rm SN1}$ and lowest $BH_{\rm RE}$ simulation are also shown for comparison in \Fig\ref{fig:comparisons_literature}.

\subsection{A unified scenario}

Although both simulations and data require further refinement, the results summarized in \Fig\ref{fig:comparisons_literature} allow us to draw several overarching conclusions. The golden mass derived from the reference \camels\ simulation, along with results from TNG100-1, TNG300-1, and the recent halo abundance matching analysis by \citet{Girelli+20}, are broadly consistent up to $z \sim 1.5$. Above this redshift, also the TNG simulations no longer show a bending in the SHMR, whereas the golden mass from \citet{Girelli+20} continues to increase with redshift, eventually overlapping with the $M_{\rm stream}$ threshold beyond $z \sim 2$.

At these high redshifts, the bending mass in the star-forming main sequence also aligns with $M_{\rm stream}$, indicating that the maximum of star formation efficiency, the main sequence bending mass, and the cold-streaming threshold all converge onto the same physical mass scale. Conversely, below $z \sim 2$, various simulations and observations---including those examining the SHMR \citep[e.g.,][]{Girelli+20}, the minimum in the DM fraction--stellar mass correlation in \camels\ and TNG, and different observations of local galaxies \citep[e.g.,][]{Tortora+19_LTGs_DM_and_slopes}---point to a characteristic mass scale, which we refer to as the golden mass, in the range $\log M_\star / M_\odot \sim 10.2$--$10.6$.

However, the bending mass of the main sequence at low redshift decreases more steeply, reaching values of $\log M_\star / M_\odot \sim 9.5$--$10$, consistent with the gas-richness threshold identified by \citet{Kannappan+13}.

In summary, above $z \sim 2$, a single characteristic mass scale appears to mark the shutdown of gas accretion, leading to quenching, morphological transformation, and the emergence of red, spheroid-dominated galaxies. Below this redshift, instead, two mass scales are emerging, one tied to the gas availability, the "gas-richness" threshold, and the second to star formation quenching, i.e. the bimodality or golden mass.

\section{Conclusions}\label{sec:conclusions}

In this paper, we investigated the emergence of the so-called "golden" mass, by studying the scaling relation between $M_{\rm DM,1/2}/M_{\star,1/2}$, the ratio of total-to-stellar mass within the half-mass radius (equivalent to a DM fraction), and stellar mass. We utilized a new suite of \camels\ simulations, which enhance previous simulations by increasing the volume to $(50\,h^{-1}\,\textrm{Mpc})^{3}$, allowing for a more detailed examination of the most massive end of the galaxy mass function (\citealt{Villaescusa-Navarro2021,Villaescusa-Navarro2022,Ni+23}). These simulations enable us to analyze the impact of various astrophysical and cosmological parameters, including SN and AGN feedback strengths as well as $\Omega_m$ and $\sigma_{8}$, across different redshifts. However, we note that larger volume simulations would be beneficial to determine whether a golden mass appears at masses greater than $10^{11}\, \rm \Msun$ at large redshifts.

For most astrophysical and cosmological parameter values across various redshifts, the $M_{\rm DM, 1/2}/M_{\star, 1/2}$--\mst\ relation follows an U-shaped curve, indicating the presence of a "golden mass" at the minimum of the curve \citep{Tortora+19_LTGs_DM_and_slopes}. However, at high redshifts, such as $z=2$, this ratio decreases with stellar mass, and no golden mass is observed.

Cosmological parameters primarily influence the normalization of this scaling relation but do not affect the emergence of the golden mass. The U-shaped trend becomes more pronounced with higher values of supernova feedback energy ($A_{\rm SN1}$) and wind velocity ($A_{\rm SN2}$), with the latter having a stronger impact on the golden mass value. Our findings confirm that while the normalization factors for AGN energy and event frequency ($A_{\rm AGN1}$ and $A_{\rm AGN2}$) have only minor or negligible effects on the scaling relations, other AGN feedback-related parameters implemented in the \camels\ simulations play a more significant role. Specifically, larger values of the normalization factor for the Bondi rate for the accretion onto BHs (BH accretion rate $BH_{\rm accr}$), the normalization factor for the limiting Eddington rate for the accretion onto BHs (Eddington factor $BH_{\rm Edd}$), the Eddington ratio, that serves as the threshold between the low-accretion and high-accretion states of AGN feedback (quasar threshold $Q_{\rm T}$), and the power-law index of the scaling of the low- to high-accretion state threshold with BH mass (quasar threshold power $Q_{\rm TP}$) increase DM fractions at high masses and induce a bend in the scaling relation, while factors such as the normalization factor for the energy in AGN feedback, per unit accretion rate, in the high-accretion state (BH feedback factor $BH_{\rm FF}$) and the fraction of the accretion rest mass that is released in the accretion process (BH radiative efficiency $BH_{\rm RE}$) work in the opposite direction.

Summarizing, the fraction of the accretion rest mass that is released in the accretion process is the most impacting parameter on the golden mass value, followed by the normalization factor for the energy in AGN feedback, the threshold between the low-accretion and high-accretion states of AGN feedback and SN wind velocity, the rest of the parameters induce milder or absent trends.

In PGs, the U-shaped curve is clearly visible in the $M_{\rm DM,1/2}/M_{\star,1/2}$--\mst\ relation, while SFGs display a monotonically decreasing trend with stellar mass across all redshifts, with only mild indications of a turnover at very low redshifts. These results align with observational findings \citep{TLBN16_IMF_dwarfs, Tortora+19_LTGs_DM_and_slopes}. In SFGs, DM fractions decrease over cosmic time, while PGs show a significant increase, particularly in the most massive systems, where size evolution contributes to the rise in central DM fractions (e.g., \citealt{Tortora+14_DMevol, Tortora+18_KiDS_DMevol}). Models with stronger SN and AGN feedback demonstrate faster evolution in both size and DM fractions.

Regarding global quantities, SFGs, which dominate the scaling relation below the golden mass, exhibit only marginal bending, which becomes more pronounced with stronger wind energy. PGs dominate at higher masses but do not show a sharp peak at the golden mass. Instead, they exhibit a mild peak around $10^{9.5-10} \, \rm M_{\odot}$ (changing to larger values for extreme values of some AGN feedback parameters) and a decreasing trend at higher masses, a pattern that warrants further observational testing. 

While there are some hints of bending in the central and global properties of SFGs, larger-volume simulations are needed to explore the massive end of the mass function and to further investigate the star formation efficiency in the most massive spiral galaxies \citep{Posti+19, Tortora+19_LTGs_DM_and_slopes}.

We have finally discussed our results in a broader context, comparing them with mass variations predicted in the shock/stream scenario, together with literature results from the stellar-to-halo mass relation and the main sequence of star-forming galaxies. Our results point to the predominant role of SN and AGN feedback in leading the scaling relations, with SN feedback impacting also the most massive galaxies. From the comparison with the literature, however, emerge some possible discrepancy on the association of the bending mass of the main sequence and the minima of our scaling relations, which possibly are two distinct mass scales (as also \citealt{Daddi+22_main_sequence} was reporting). We believe that the bending mass in the main sequence is equivalent to the "gas-richness" mass mentioned in \cite{Kannappan+13}, separating "accretion-dominated" by "processing dominated" galaxies, while the minima in our scaling relations are more connected to morpholodical differences and a quenching of star formation, more than on the availability of gas. We speculate that above $z \sim 2$, all these characteristic mass scales converge, likely coinciding with the cold-streaming threshold mass ($M_{\rm stream}$). Below this redshift, however, two distinct mass scales emerge: the gas-richness threshold at $\sim 10^{9.5\text{--}10} \, \Msun$, and the bimodality or golden mass at $\sim 10^{10.2\text{--}10.6} \, \Msun$.

In future papers, following the procedure in Paper I and II, we will constrain these new simulations with such extended list of simulation parameters with observations (star forming galaxies: \citealt{Tortora+19_LTGs_DM_and_slopes}; passive galaxies: \citealt{SPIDER-VI}; \citealt{Zhu2023}), weighting the role of SN and AGN feedback. Among the new data available, the Euclid mission's $>$100,000 strong lenses (\citealt{Mellier+24_EuclidI,Acevedo-Barroso+24_ELSE}), combined with weak lensing signals from stacked lenses and spectroscopic follow-ups (e.g. \citealt{Collett+23_4SLSLS}), will provide precise data on central DM and total virial mass across a broad mass range and up to $z=2$, allowing to constrain physical processes, and in particular AGN feedback (e.g. \citealt{Mukherjee19_SEAGLEII}), which is one of the most dominant phenomena in very massive galaxies. Moreover, discovering several systems with spiral galaxies as lenses will also help address questions about the star formation efficiency of the most massive spiral galaxies across cosmic time, constraining the golden mass as a function of time and galaxy type.

\begin{acknowledgements}
We thank the referee for their insightful comments, which have significantly enhanced the clarity and quality of our manuscript. C. T. thanks Carlo Cannarozzo for the useful discussions during IAU Symposium \#396. C.T. and V.B. acknowledge the INAF grant 2022 LEMON.
\end{acknowledgements}

\bibliography{paper_v2}

%myrefs

%\begin{appendix}
%\end{appendix}

\appendix

\section{Star-forming main sequence compared with the literature}\label{app:SFRmass_literature}

In this paper we distinguished SFGs from PGs adopting a threshold of $\log(\textrm{sSFR}/\textrm{yr}^{-1}) = -10.5$. However, measuring the SFR is also essential to study one of the most widely explored scaling relations in galaxy evolution: the star-forming main sequence \citep{Speagle+14,Whitaker+14,Lee+15,Ginolfi+20_MAGMAI,delvecchio+21,Daddi+22_main_sequence,Popesso+23}.

\begin{figure*}
\centering
\includegraphics[width=\linewidth]{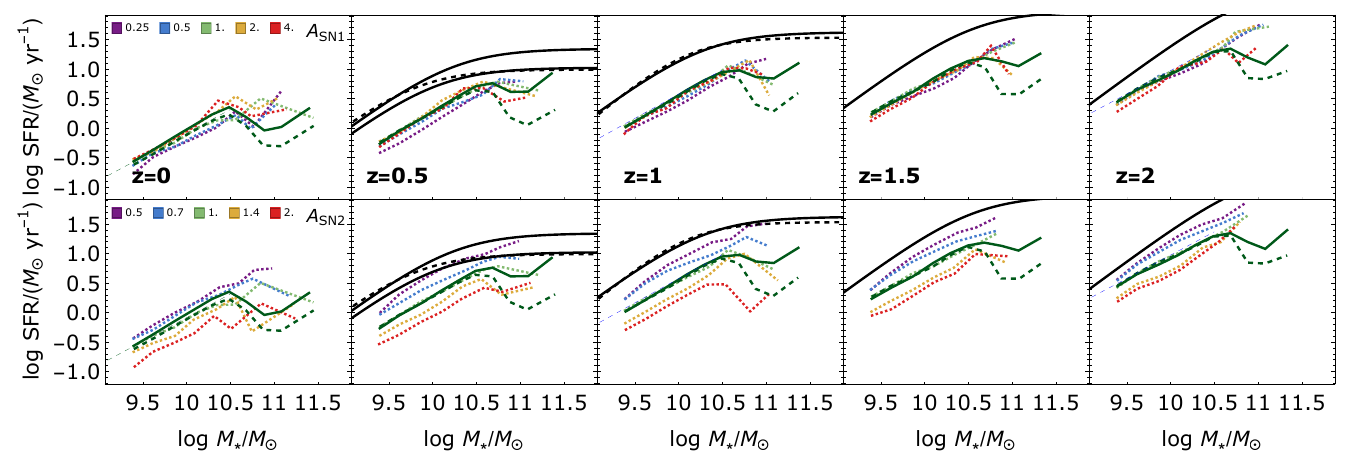}
\caption{The star-forming main sequence in the \camels\ simulations is shown as a function of $A_{\rm SN1}$ (top panels) and $A_{\rm SN2}$ (bottom panels). To illustrate the redshift evolution, we present simulation snapshots at $z=0$, $z=0.5$, $z=1$, $z=1.5$, and $z=2$, displayed from left to right. In each panel, we show the median SFR in stellar mass bins.
We compare the \camels\ results with those from the larger-volume TNG100-1 (solid dark green) and TNG300-1 (dashed dark green) simulations \citep{Nelson2019a}. The thin dashed blue line represents the best-fit to the main sequence in the stellar mass range $\log(\mst/\Msun) = 9$–$10.5$, as derived in \cite{Donnari+19_TNG} using TNG300-1. We also include observational results from \citet[][solid black line]{delvecchio+21} and \citet[][dashed black line]{Lee+15} in contiguous redshift bins (extracted from Table 1 in \citealt{Daddi+22_main_sequence}).}
\label{fig:fig_SFR_Mstar_SNfeed}
\end{figure*}

In \Fig\ref{fig:fig_SFR_Mstar_SNfeed}, we show the star-forming main sequence for \camels\ galaxies. Differently from the rest of the paper, here we adopt a more conservative threshold of $\log(\textrm{sSFR}/\textrm{yr}^{-1}) = -12$ to select SFGs. This choice reflects the fact that the main sequence in TNG simulations shows curvature depending on the specific sSFR cut adopted (e.g., \citealt{Donnari+19_TNG}). Nonetheless, applying the same selection used in this section to the other main scaling relations in the paper does not affect our main conclusions.

We find that the energy injected by SN winds ($A_{\rm SN1}$) has a smaller impact on the normalization of the main sequence compared to wind velocity ($A_{\rm SN2}$). At fixed stellar mass, higher wind velocities result in lower SFRs. Moreover, using the same sSFR threshold as in the rest of the paper tends to wash out the bending of the main sequence.

Due to the limited volume of our \camels\ boxes, we cannot probe the high-mass end where the bending becomes more prominent. To mitigate this, we compare our results with the larger-volume TNG100-1 and TNG300-1 simulations, which cover volumes 3.4 and 69 times larger, respectively, than the volume used in our \camels\ simulations. TNG100-1 offers higher mass and spatial resolution, while TNG300-1 has a resolution comparable to \camels. The \camels\ reference simulation is in excellent agreement with these runs at intermediate masses. At higher masses, some discrepancies emerge, likely driven by low-number statistics. We also include in the $z=0$, $z=1$, and $z=2$ panels the best-fit relations from \cite{Donnari+19_TNG}, derived for galaxies in the stellar mass range $\log \mst/\Msun \in (9.0$–$10.5)$ in TNG300-1. Despite minor differences in the selection criteria, our trends in TNG300-1 match theirs very well. 

Finally, we compare our results with empirical fits from \cite{Lee+15} and \cite{delvecchio+21}. These confirm that both the slope and the redshift evolution of the simulated main sequence are broadly consistent with observations. However, the TNG simulations systematically underpredict the SFR of real galaxies by about 0.5 dex. Interestingly, we find that adopting very low values of $A_{\rm SN2}$ (e.g., $\sim 0.5$) can help reduce this discrepancy. Our results follow the findings in \cite{Donnari+19_TNG}. For a detailed analysis of the main sequence and the quenched fraction in TNG simulations, we refer the reader to that paper (see also \citealt{Hahn+19}).

\section{The stellar-to-halo-mass relation compared with the literature}\label{app:SGMR_literature}

The SHMR is one of the most extensively studied scaling relations in the literature. Results for \camels\ simulations, dividing galaxies in PGs and SFGs, are shown in \Fig\ref{fig:fstar_vs_Mstar_types}. It is however important to compare our results with existing findings.

\begin{figure*}
\centering
\includegraphics[width=\linewidth]{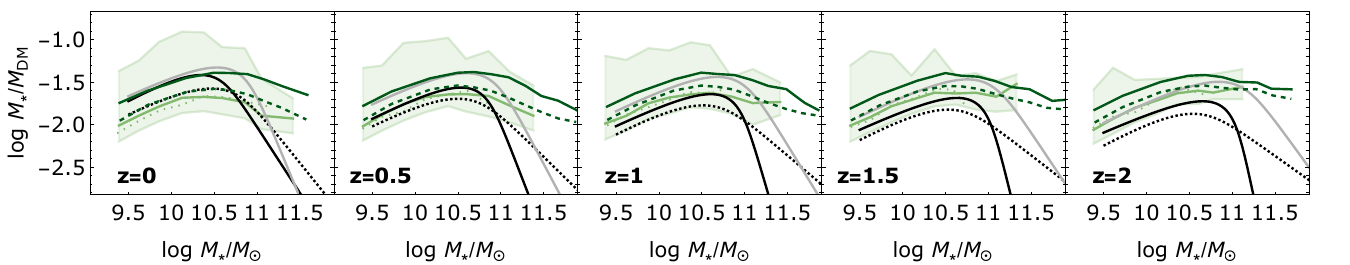}
\caption{SHMR of the reference \camels\ simulation with a volume of $(50\, \textrm{h}^{-1}\,\textrm{Mpc})^{3}$ showing the medians in stellar bins (solid green line) and the 16th–84th percentile range (shaded green region). To examine the evolution across redshift, simulation snapshots at $z=0$, $z=0.5$, $z=1$, $z=1.5$ and $z=2$ are displayed from left to right. It is compared with the same simulation run in a smaller volume of $(25\, \textrm{h}^{-1}\,\textrm{Mpc})^{3}$ \citep[][dotted green]{Villaescusa-Navarro2021}. We also show results from the TNG100-1 (dark green) and TNG300-1 (dark dashed green) simulations \citep{Nelson2019a}. For both \camels\ and TNG simulations, we plot medians of the sample distribution divided in stellar mass bins. For comparison with semi-empirical models, we include abundance matching results from \citet[][dashed black]{Moster+10}, \citet[][solid black]{Moster2013}, and \citet[][Table 3, solid gray]{Girelli+20}.}
\label{fig:fig_SHMR_literature}
\end{figure*}

As a first step, in \Fig\ref{fig:fig_SHMR_literature} we compare our results with those obtained using a \camels\ simulation volume of $(25\,\textrm{h}^{-1}\,\textrm{Mpc})^{3}$. This comparison shows that the SHMR remains largely unchanged; however, the smaller volume contains fewer galaxies, which affects the relation, especially at the high-mass end.

We also compared our results with those from the TNG100-1 and TNG300-1 simulations. At fixed stellar mass, TNG100-1 predicts higher halo masses (by $\sim 0.2$–$0.3$ dex), whereas the SHMR from TNG300-1 more closely matches that from \camels. The benefits of the larger simulation volumes in TNG100-1 and TNG300-1 are evident, highlighting a clear path forward for accurately mapping this and other scaling relations, particularly at the high-mass end. 

Finally, we compared our results with semi-empirical models based on halo abundance matching. We invert Equation 2 in \cite{Moster+10} to obtain the correlation between halo and stellar mass, and use best-fitted parameters in Table 7 of \cite{Moster+10}, Table 1 of \cite{Moster2013} and Table 3 of \cite{Girelli+20} and the respective functional forms for the evolving parameters with redshifts. The models of \citet{Moster+10} and \citet{Moster2013} predict stronger evolution at $z < 0.5$ than seen in both \camels\ and TNG, especially at stellar masses $\lesssim 10^{10.5}\, \rm \Msun$. At $z \sim 0$, \cite{Moster2013} resembles TNG100-1, while at higher redshift they align more closely with \camels. \cite{Moster+10} is in better agreement with TNG300-1 and \camels. In contrast, the results from \citet{Girelli+20} closely follow TNG100-1 across all redshifts up to $\lesssim 10^{10.5}\, \rm \Msun$. Notably, all abundance matching models exhibit a steep decline at the high-mass end, a trend not seen in the TNG and \camels\ simulations, which instead show a more gradual decrease beyond $\sim 10^{10.5}\, \rm \Msun$.

\end{document}